\documentclass[superscriptaddress,showpacs,twocolumn,prl]{revtex4}

\usepackage{amsmath}
\usepackage{amssymb}
\usepackage{dsfont}
\usepackage{color}
\usepackage{hyperref}
\usepackage{graphicx}

\newcommand*{\balancecolsandclearpage}{
  \clearpage
  \twocolumngrid
}

\definecolor{Nathanblue}{rgb}{0.,0.24,0.51}

\begin{document}

\title{Observation of a localized flat-band state in a photonic Lieb lattice}
\author{Sebabrata Mukherjee}
\email{snm32@hw.ac.uk}
\affiliation{SUPA, Institute of Photonics and Quantum Sciences, Heriot-Watt University, Edinburgh, EH14 4AS, United Kingdom}
\author{Alexander Spracklen}
\affiliation{SUPA, Institute of Photonics and Quantum Sciences, Heriot-Watt University, Edinburgh, EH14 4AS, United Kingdom}
\author{Debaditya Choudhury}
\affiliation{SUPA, Institute of Photonics and Quantum Sciences, Heriot-Watt University, Edinburgh, EH14 4AS, United Kingdom}
\author{Nathan Goldman}
\affiliation{ Center for Nonlinear Phenomena and Complex Systems, Universit\'e Libre de Bruxelles, CP 231, Campus Plaine, B-1050 Brussels, Belgium}
\affiliation{ Laboratoire Kastler Brossel, Coll\`ege de France, 11 place Marcelin Berthelot, 75005, Paris, France}
\author{Patrik \"Ohberg}
\affiliation{SUPA, Institute of Photonics and Quantum Sciences, Heriot-Watt University, Edinburgh, EH14 4AS, United Kingdom}
\author{Erika Andersson}
\affiliation{SUPA, Institute of Photonics and Quantum Sciences, Heriot-Watt University, Edinburgh, EH14 4AS, United Kingdom}
\author{Robert R. Thomson}
\affiliation{SUPA, Institute of Photonics and Quantum Sciences, Heriot-Watt University, Edinburgh, EH14 4AS, United Kingdom}

\begin{abstract}
{We demonstrate the first experimental realization of a dispersionless state, in a photonic Lieb lattice formed by an array of optical waveguides. This engineered lattice supports three energy bands, including a perfectly flat middle band with an infinite effective mass. We analyse, both experimentally and theoretically, the evolution of well-prepared flat-band states, and show their remarkable robustness, even in the presence of disorder. The realization 
of flat-band states in photonic lattices opens an exciting door towards quantum simulation of flat-band models in a highly controllable environment.}
\end{abstract}

\maketitle

{ {\it Introduction.}  Transport in crystals reveals a rich variety of phenomena, ranging from dissipationless currents in superconductors to spin-polarized edge-states in topological insulators. Importantly, both classical and quantum transport can 
show localization effects, which typically depend on the dimensionality, 
presence of disorder or impurities, and 
nature of 
inter-particle interactions. Localization phenomena include disorder-induced (Anderson) localization, which is now well-established in non-interacting systems~\cite{Billy:2008kd}, and also many-body localization, as recently explored in disordered cold atomic gases \cite{schreiber2015}. Interestingly, localization can also exist in lattices 
without disorder. Indeed, 
specific lattice geometries can 
allow for destructive wave interference, leading to perfectly flat (dispersionless) energy bands where particles 
exhibit infinite effective mass. Perhaps the simplest lattice where this phenomenon occurs is the two-dimensional Lieb lattice, Fig. \ref{fig1} (a),
which belongs to a wide family of flat-band models \cite{Mielke:1992ct,Aoki:1996ik,Deng2003,Wu:2007iz,Tasaki2008,Lan:2012gz,Jacqmin:2014cj}. Originally, flat-band Hubbard models were analyzed in the context of magnetism, where electrons populating the flat band were found to contribute to unusual ferromagnetic ground states~\cite{Tasaki2008}. More recently, the interplay between flat-band localization and correlated disorder was studied in Ref. \cite{bodyfelt2014}.

It is intriguing that slight changes in the tunneling matrix elements between different lattice sites can lead to dramatically different transport properties, including exact localization associated with flat-band states.  Different physical platforms, including cold atoms in optical lattices  \cite{Shen:2010jz,Apaja2010,Goldman2011} and light propagation in photonic crystals \cite{Vicencio2014,Guzman-Silva2014}, have been envisaged to reveal flat-band properties through the engineering of exotic 
specific lattice models. Both cold atoms and photonic crystals offer a high control over the lattice geometry, and  allow for the addition of tunable disorder or interactions \cite{Bloch:2008gl,Carusotto:2013gh}. This suggests an exciting route for the quantum simulation of interacting flat-band systems. Recently, Guzman-Silva et. al. \cite{Guzman-Silva2014} have reported  bulk and edge transport phenomena in a photonic Lieb lattice. However, to date, diffraction-free propagation of a flat-band state has not been observed. In this Letter, we present the first experimental observation of a stationary and localized flat-band state in a photonic Lieb lattice, where the lattice is formed by a 
two-dimensional array of optical waveguides, fabricated using femtosecond laser writing~\cite{Davis1996}.

\begin{figure}[h!]
\centering
\includegraphics[width=8.5cm]{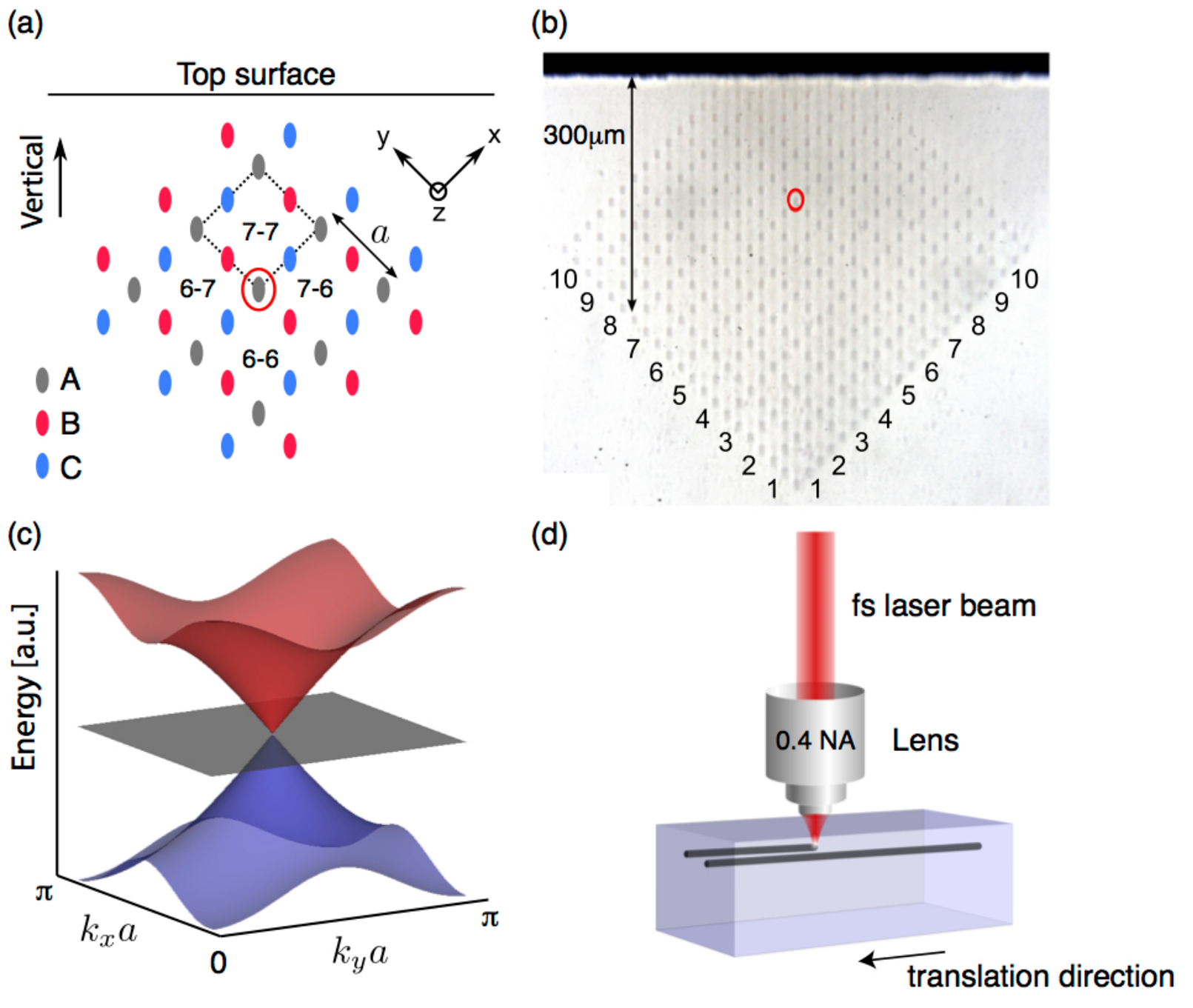}
\caption{(a) Edge-centered square (Lieb) lattice. The basis contains three sites (A, B and C). 
To avoid edge effects and effects due to lattice inhomogeneity with depth,  all measurements were performed near the circled A-site ($A_{7-7}$), see also (b).  
The lattice constant {{$a$ = 44 $\mu m$}}. (b) White-light transmission optical micrograph of the facet of a Lieb lattice with 323 waveguides fabricated by femtosecond laser writing. Each waveguide supports only a single fundamental mode at 780 nm. The  next-nearest-neighbor coupling for a 7 cm long glass chip was observed to be negligible. To minimize the difference in the  next-nearest-neighbor coupling constants along the $x$- and $y$-axes, the lattice was {{ fabricated such that the $x$- and $y$- axes of the lattice were at 45$^{\circ}$ relative to the vertical axis}}
(c) Representation of the three energy bands, including the flat band in the middle, for $\{k_xa,k_ya\}=[0,\pi]$. (d) The femtosecond laser writing technique.}
\label{fig1}
\end{figure}

\begin{figure}[!]
\includegraphics[width=1.0\linewidth]{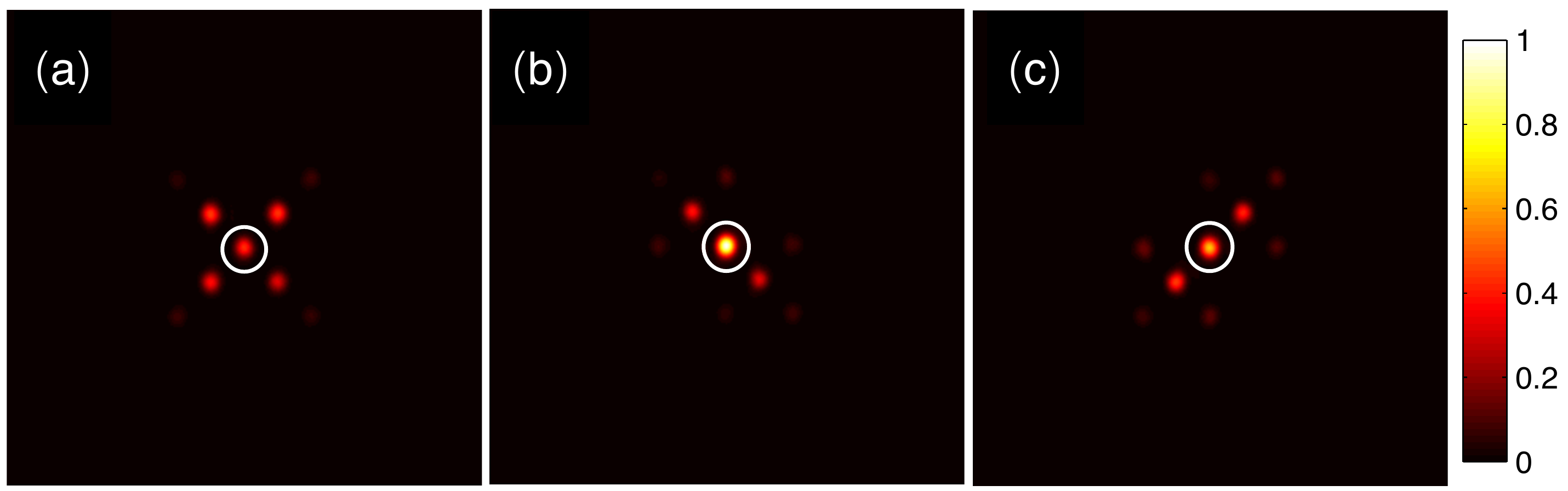}
\caption{Average diffraction patterns when A (a), B (b) and C (c) sites surrounding $A_{7-7}$ were excited separately. The excited sites are circled. Each image is normalized such that the total output power is 1, and the field of view is approximately 225 $\mu m$ by 225 $\mu m$. Simulations based on these averaged diffraction patterns indicate a waveguide-to-waveguide coupling constant of 0.01 $\pm$0.001 $mm^{-1}$.}
\label{fig2}
\end{figure}

Light propagation
in an array of evanescently coupled optical waveguides, i.e. a photonic lattice, is in the paraxial approximation described by a 
Schr\"odinger equation 
\begin{equation}
i\partial_z\Psi(x,y,z)=\left[-\frac{1}{2k_0n_0}\nabla_\perp^2-k_0\Delta n(x,y,z)\right]\Psi(x,y,z) \label{schr}
\end{equation}
where the refractive index profile across the lattice ($\Delta n(x,y,z)$) acts as an effective potential for the light field. The role of the wavefunction is played by the envelope of the electric field $E(x,y,z)=\Psi(x,y,z) e^{i(k_0z-\omega t)}$, where $k_0$ is the free-space wavenumber and  $n_0$ is the refractive index of the host material the lattice is created in. By 
controlling the refractive index ($\Delta n$) 
profile across the lattice structure, 
one may use photonic lattices to observe and probe phenomena from 
solid-state physics, such as Bloch oscillations \cite{Pertsch1999, Morandotti1999, Lenz1999, Chiodo2006}, dynamic localization \cite{Dreisow2008}, Bloch-Zener oscillations \cite{Dreisow2009}, and Landau-Zener tunneling  \cite{2Dreisow2009}.

For weak evanescent coupling, Eq. (\ref{schr}) can be modeled by a tight-binding Hamiltonian. In this paper we consider an edge-centered square (Lieb) lattice, as shown in Fig. \ref{fig1} (a). 
Fourier transforming the Hamiltonian into $k$-space, gives an energy spectrum with three bands,
\begin{eqnarray}
\Omega_\pm (\textbf{k})&=&\pm2\sqrt{\kappa_x^2 \cos^2(k_x a)+\kappa_y^2 \cos^2(k_y a)} \\
\Omega_0 (\textbf{k})&=&0,
\end{eqnarray}
where $\kappa_x$ and $\kappa_y$ are the hopping amplitudes (coupling constants)  for nearest-neighbor sites along the  $x$- and $y$-axes, and $a$ is the lattice constant. $\Omega_\pm$ are the energies of the upper and the lower bands, respectively, and $\Omega_0$ represents the non-dispersive flat band. The Brillouin zone spans $0<k_x,k_y<\pi/a$. The three bands intersect at $k_x=k_y=\pi/2a$, known as the M point, see Fig. \ref{fig1}(c).}


{The lattice Hamiltonian displays particle-hole symmetry.} 
This symmetry, combined with the statement that at each $\boldsymbol{k}$ there are three energy states, automatically implies a flat band, as for each $\boldsymbol{k}$, one of these energies must be zero. This argument breaks down in presence of disorder, as $k_{x,y}$ are no longer good quantum numbers. However, as shown in  the Supplementary material \cite{sup}, the flat band persists for  
off-diagonal disorder (i.e. disordered coupling constants)  
of  arbitrary strength. This form of disorder is  
present in our lattices, and is due to small random variations in waveguide-to-waveguide separations across and along the lattice. In contrast, diagonal disorder would occur if different waveguides exhibited random variations in their propagation constants. As we 
discuss
later, this form of disorder is not significant in our lattices.

The dotted square in Fig. \ref{fig1} (a) shows a primitive cell of the lattice. There are four A-sites at the corners of each cell; two B-sites and two C-sites lie on the edges.
If the Lieb lattice is isotropic, with  $\kappa_x=\kappa_y$, then a superposition of states in the flat band can be excited if (a) the lattice has insignificant next-nearest-neighbor coupling, and (b) the two B-sites and two C-sites of a primitive cell are excited with equal intensities ($I_B=I_C$) and alternating phases ($\phi_B=\phi_C\pm\pi$).  In this Letter, we demonstrate experimentally and theoretically that the flat-band state excited at the input of a photonic Lieb lattice remains localized and does not diffract.

{\it Fabrication of photonic Lieb lattice.} Photonic Lieb lattices were fabricated using  femtosecond laser writing, a well-established laser fabrication technique~\cite{Davis1996}. 
The substrate material (Corning Eagle$^{2000}$) was mounted on air-bearing Aerotech  $x$-$y$-$z$ translation stages (ABL1000), and each lattice waveguide was fabricated by translating the substrate once through the focus of a 500 kHz train of circularly polarized sub-picosecond ($\sim$400 fs) laser pulses, generated by a Menlo BlueCut fibre laser system. The 
laser writing parameters were optimized to produce low propagation loss, single-mode waveguides for operation at 
a wavelength of 780 nm. The waveguide refractive index profile 
was controlled using the  slit-beam shaping method \cite{Ams2005, Cheng2003}, by placing a slit directly in front of the 0.4 numerical aperture (NA) lens used to focus the laser pulses inside the substrate. The effective NA's of the laser focus were calculated to be $\approx$ 0.2 and 0.3,  along the axis perpendicular and parallel to the waveguide axis, respectively. The final Lieb lattices were inscribed in a 7 cm long glass chip. Individual waveguides exhibited propagation loss of $\approx$1 dB/cm at 780 nm.

\begin{figure}[h!]
\includegraphics[width=1.0\linewidth]{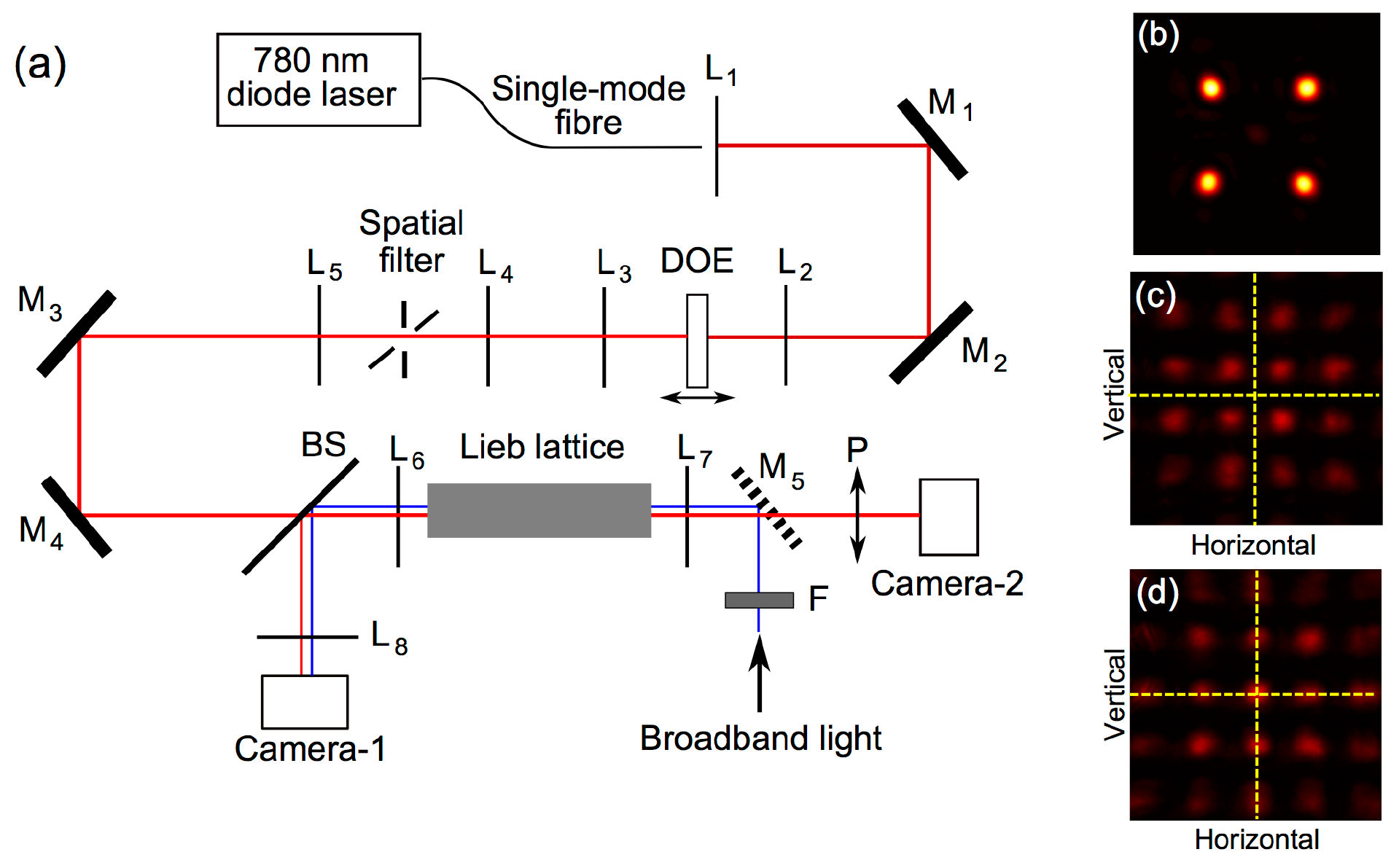}
\caption{{{(a) Experimental set-up for exciting the flat band. 
L$_1$-L$_8$ are convex lenses, M$_1$-M$_5$ are silver-coated mirrors with M$_5$  
on a flipped mount, and BS is a beam splitter.
L$_1$ collimated the 780 nm light emerging from a single-mode fibre. A zero-order nulled, binary-phase, square-checker-board diffractive optical element (DOE) generated a square array of diffraction-order ``spots'' at the focus of L$_2$. Using L$_3$ and L$_4$, these spots were relay-imaged to the spatial filter, the transmission aperture of which was adjusted to pass only the four first orders (and a very weak 0$\textsuperscript{th}$ order). Using L$_5$ and L$_6$, the four spots were relay-imaged to the input facet of the Lieb lattice. The output facet of the lattice was flood-illuminated using 780$\pm$10 nm light, filtered from a broadband white-light source using a bandpass filter (F). This flood-illumination excited the lattice modes, and enabled precise alignment of the the four spots with the desired waveguide modes using Camera-1. Once light had been coupled to the lattice, the light distribution at the output facet could be viewed using Camera-2. A polarizer (P) passing only vertically polarized light was placed in front of Camera-2, ensuring that measurements were not affected by polarization-dependent coupling in the lattice. 
(b) Intensity profile of the four spots which were coupled into the lattice (field of view approximately 80 $\mu m$ by 80 $\mu m$). Note the very weak 0$\textsuperscript{th}$ order. (c $\&$ d) Fraunhofer diffraction patterns for the flat-band state and equal-phase state respectively. The yellow dotted lines represent the positions of the optic axis.}
}}
\label{fig3}
\end{figure}

\begin{figure}[h!]
\includegraphics[width=1.0\linewidth]{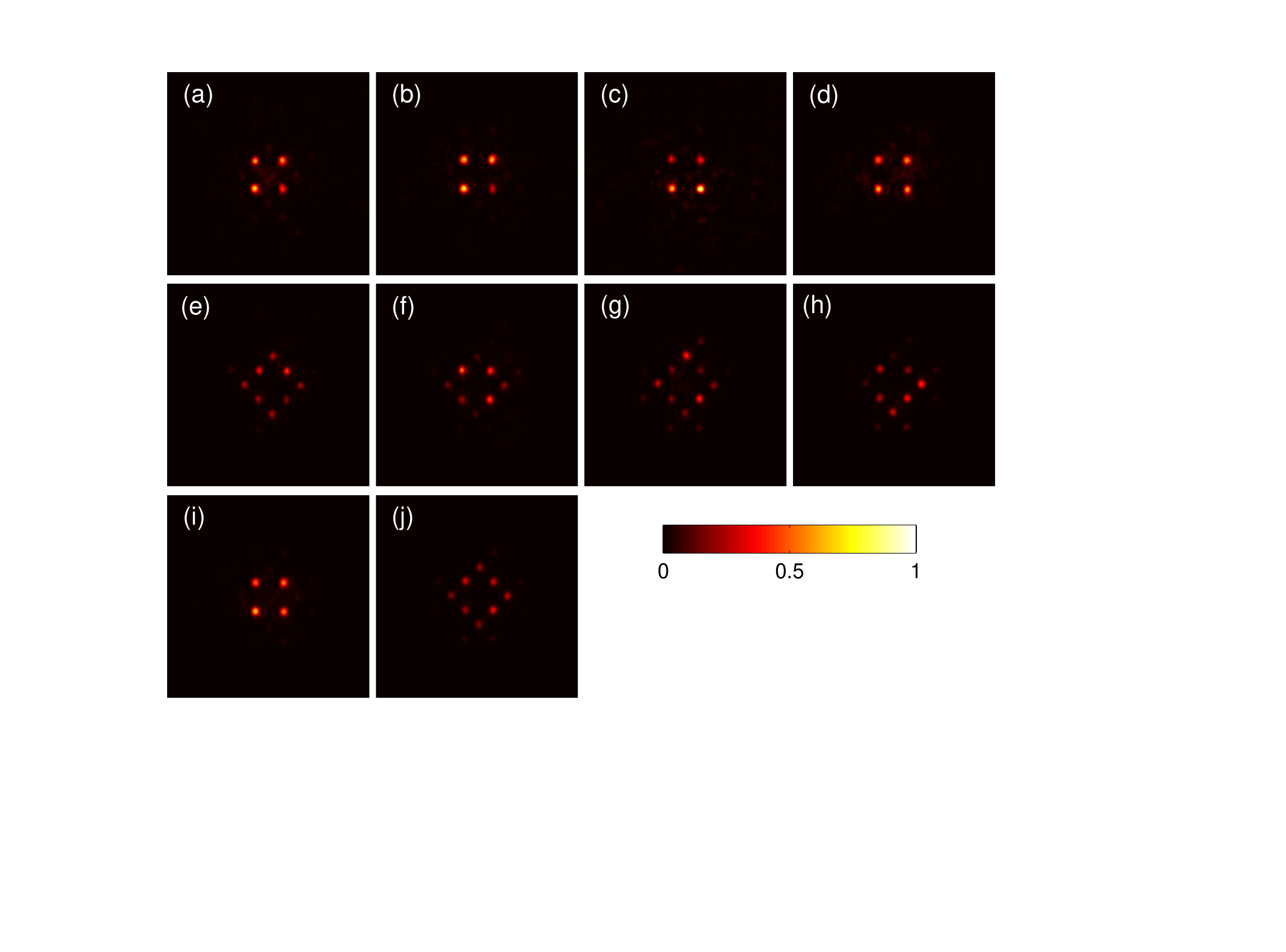}
\caption{(a-d) are the non-diffracting states observed at the output of the lattice when the flat-band state was launched into the B and C sites of the (7-7), (6-7), (7-6) and (6-6) cells respectively. When the equal-phase state was launched into the same cells, the output state is not localized, as shown by (e-h).  (i) shows the average diffraction pattern of (a-d) and (j) is the average diffraction pattern of (e-h). Each image is normalized so that total intensity is 1. The field of view is approximately 210 $\mu m$ by 210 $\mu m$.}
\label{fig4}
\end{figure}

Thirteen complete Lieb lattices (lattice constant $a$ = 24 to 48 $\mu m$ in steps of 2 $\mu m$) were fabricated, each containing 323 single-mode waveguides. However, as discussed later, it was not possible to observe the non-diffracting state when $a\le$ 42 $\mu m$,
because next-nearest-neighbor coupling in these lattices is non-negligible over the 7 cm length of the chip, destroying the flat band. The remainder of the paper shall, therefore, focus on the lattice fabricated  with $a$ = 44 $\mu m$ -- the most compact Lieb lattice fabricated where the flat band could be excited. 
A white-light transmission optical micrograph of the facet of this lattice is shown in Fig. 1 (b). The individual waveguide modes are slightly elliptical in shape, 
supporting a single mode at 780 nm with major and minor mode-field diameters of 8.6 and 7.4 $\mu m$ along the vertical and horizontal axis respectively.
To minimize differences in nearest-neighbor coupling coefficients along the $x-$ and $y-$axes, the lattice was fabricated such that the $x-$ and $y-$axes were at 45$^{\circ}$ relative to the vertical and horizontal axis, Fig. \ref{fig1}(b).

{{
{\it Optical characterization and excitation of photonic Lieb lattice.}
It is well known that for femtosecond laser writing, depth-dependent aberrations imparted on the laser beam by the air-glass interface can significantly affect the properties of the written waveguides \cite{Salter2014}. This will clearly result in depth-dependent lattice properties. To assess the homogeneity of the fabricated Lieb lattice, we investigated how the coupling constant between two evanescently coupled waveguides varied as a function of depth. We inscribed arrays of two-waveguide evanescent field couplers at six different depths, from 100 to 600 $\mu m$ in steps of 100 $\mu m$. Each coupler was fabricated using the same waveguide-to-waveguide separation in the interaction region (22 $\mu m$), and same waveguide-to-waveguide angle (relative to the vertical axis in Fig. \ref{fig1} (a)) as the waveguides in the Lieb lattice. At each depth, 5 sets of 17 couplers were fabricated, each set consisting of couplers with interaction lengths between 1 and 65 mm, in steps of 4 mm. Each coupler was characterized by injecting 780 nm light into one of the waveguides and measuring the output coupling ratio. Using this data, and the procedure outlined in \cite{Eaton2009}, the mean and standard deviation of the coupling constant at each depth was evaluated to be $\approx$ 0.01 mm$^{-1}$ and $\approx$ 0.002 mm$^{-1}$ respectively for couplers fabricated up to a maximum depth of 300 $\mu m$. After this, both the coupling constant and variance were observed to become a function of depth, with deeper structures exhibiting a progressively higher variance.

To investigate whether the observed variation in coupling constant was due to random variations in the waveguide-to-waveguide separation (off-diagonal disorder), or random variations in waveguide propagation constants (diagonal disorder), we performed a second set of experiments, where an array of couplers with different interaction lengths were fabricated at a single depth inside the substrate. 
The waveguide-to-waveguide angle and separation in the interaction region were set to 45$^{\circ}$ and 15 $\mu m$ respectively -- the reduced interaction separation was used to
reduce the coupling length. Coupling characteristics for these couplers were observed to be close to ideal, with near-complete transfer of power from input waveguide to the other waveguide achieved after one coupling length. 
Since complete transfer of energy from one waveguide to the another in an evanescent field coupler is only possible if the waveguides support modes with identical propagation constants, we conclude that local variations in waveguide propagation constants are negligible in our Lieb lattices. Hence, diagonal disorder is not significant.
}}


Given the results outlined above, all optical measurements of the Lieb lattice were performed by injecting light into the primitive sites surrounding the $A_{7-7}$ site (circled in Fig. \ref{fig1} (b)), which is 
at a depth of $\approx$150 $\mu m$. First, 780 nm light was individually coupled to the nine A-, six B- and six  C-sites surrounding the $A_{7-7}$ site, and the output diffraction patterns were measured. For each type of injection site, the obtained diffraction patterns were normalized and averaged. The results of these measurements are presented in Fig. \ref{fig2}, where the sites excited at the input are circled. It can clearly be seen that the circled A-site in Fig. \ref{fig2} (a) contains less light than the circled B-site in Fig. \ref{fig2} (b), or circled C-site in Fig. \ref{fig2} (c), confirming that light injected into A-sites diffracts more than light injected into B- or C-sites. This has been recently shown by Guzman-Silva et. al. \cite{Guzman-Silva2014}. 
 
Fig. \ref{fig3} (a) shows a schematic of the experimental set-up used to excite the flat-band state. 
Laser light of 780 nm emerging from a single-mode fibre was collimated using lens L$_1$. This light was then focused through a zero-order nulled (for 780 nm) diffractive optical element (DOE) (binary-phase, square-checker-board pattern) using lens L$_2$, to generated a square array of diffracted orders at the focus of L$_2$. Using lenses L$_3$ and L$_4$, these diffraction orders were relay-imaged to the spatial filter, which blocks all orders, except for the four first-order ``spots''. Using lenses L$_5$ and L$_6$, the four spots were relay-imaged to the input-facet of the Lieb lattice (Fig. \ref{fig3}(b)). The size of each spot on the lattice facet could be controlled via the diameter of the beam entering lens L$_6$ and its focal length, and the spacing between the spots could be controlled via the distance of the DOE from lens L$_2$. The relative optical phases between the spots could be inferred by viewing the four-spot interference pattern in the Fraunhofer regime (using a camera not shown). 
 As shown in Figs. \ref{fig3} (c) and (d), the relative phases between the spots could be controlled by translating the DOE in the $x$-$y$-plane (the $z$ axis is the beam propagation axis). Independent control over the spacing and size of the four spots on the facet of the Lieb lattice enabled simultaneous excitation of two B- and two C-site waveguides for a given primitive cell. To couple the spots to the lattice in a controllable manner, 780$\pm$10 nm light from a filtered broad-band source was used to flood-illuminate the output end of the lattice and excite all the guided modes. Using Camera-1, it was thus possible to simultaneously view both the lattice modes and the excitation spots, enabling us to launch light specifically to the modes of the lattice. 

To excite the flat-band state, the DOE position was set to produce four equal intensity spots (relative standard deviation (RSD): 5.7 $\%$), with alternating $0$ and $\pi$ phases, Fig. \ref{fig3} (c). These four spots were then coupled to the B- and C-sites for the chosen 
primitive cell. A proof that such a state will excite the flat band is given in the Supplementary material. The (7-7), (6-7), (7-6) and (6-6) primitive cells were each excited, 
and the output diffraction patterns observed using Camera-2. As shown in Fig. \ref{fig4} (a-d), when the flat-band state was excited, no significant tunneling of light into the surrounding lattice sites could be observed after 7 cm of propagation. Non-diffracting states remain localized. Absence of diffraction was not observed in lattices with $a\le$ 42 $\mu m$, which we attribute to non-negligible next-nearest-neighbor coupling.

To confirm that the observed absence of diffraction
is particular to the phase and intensity distribution of the injected state, we coupled another state to the lattice, where B- and C-sites of a primitive cell are excited with close to equal intensity (RSD: 5.7 $\%$) but equal phase (the equal-phase state), Fig. \ref{fig3} (d). As presented in Fig. \ref{fig4} (e-h), this state is not localized when injected into the (7-7), (6-7), (7-6) and (6-6) primitive cells, due to its orthogonality to the flat-band-state. Interestingly, the diffraction patterns shown in Figs. \ref{fig4} (e-h) are different, 
depending on which cell is excited, due to off-diagonal disorder. As discussed earlier, we are still able to successfully excite the flat band (see also Supplementary material \cite{sup} for a theoretical description). 

{\it Conclusions.} We have experimentally  
excited a flat-band state in a photonic Lieb lattice, and observed non-diffractive propagation. Such states may provide useful applications  in, for instance, image processing and precision measurements. It 
would be intriguing to extend this study to the case where nonlinearities are present. 
This would provide a platform to simulate and investigate the behavior of interacting particles with flat dispersion relations, suggesting an interesting route towards strongly-correlated states of matter in photonic systems \cite{Umucallar:2012bo,Maghrebi:2014tq}. 

While in the final stage of preparing the current manuscript we became aware of similar work by Vicencio et al.  \cite{Vicencio2014b}.

\begin{acknowledgments} R.R.T. gratefully acknowledges funding from the UK Science and Technology Facilities Council (STFC) in the form of an STFC Advanced Fellowship (ST/H005595/1) and through the STFC Project Research and Development (STFC-PRD) scheme (ST/K00235X/1). RRT also thanks the European Union for funding via the OPTICON Research Infrastructure for Optical/IR astronomy (EU-FP7 226604). S.M. and R.R.T. thank Andrew Waddie and Neil Ross for designing and fabricating the DOE respectively. A.S. acknowledges support from the EPSRC CM-DTC. S.M. thanks Heriot Watt University for a James-Watt Ph.D Scholarship. N.G. is financed by the FRS-FNRS Belgium. We acknowledge helpful discussions with Manuel Valiente.
\end{acknowledgments}



\balancecolsandclearpage

{{\bf SUPPLEMENTARY MATERIAL}}
\vskip 0.5cm

In this supplementary material, we first show that the non-diffracting input state consists of only flat-band eigenstates when the horizontal and vertical couplings are equal. Next we consider the effect of disorder, and show that the flat band persists also in the presence of off-diagonal disorder. This type of disorder is known to be present in the experiment, as the coupling strengths between waveguides is highly sensitive to their spacing.

\section{1. The non-diffracting input state lies in the flat band}

The tight-binding Hamiltonian for the Lieb Lattice can be written as
\begin{eqnarray}
\hat H&=&\sum_{(n,m)}(\kappa_y \hat b_{n,m}^{\dagger} \hat a_{n,m}+\kappa_y \hat b_{n,m-1}^{\dagger}\hat a_{n,m}+\nonumber\\
&&\kappa_x \hat c_{n,m}^{\dagger} \hat a_{n,m}+\kappa_x \hat c_{n-1,m}^{\dagger} \hat a_{n,m})+\text{h.c.},
\end{eqnarray}
where $\hat a_{n,m}$, $\hat b_{n,m}$ and $\hat c_{n,m}$ are the destruction operators for the A, B and C sites in lattice unit $(n,m)$, respectively.  
Fourier transforming the real-space tight-binding Hamiltonian results in a $k$-space Hamiltonian which is a 3x3 matrix due to the three inequivalent lattice sites per unit cell,
\begin{eqnarray} 
\hat H &=&\sum_{\textbf{k}} 
\begin{pmatrix}
  \hat a^{\dagger}_\textbf{k} & \hat b^{\dagger}_\textbf{k} & \hat c^{\dagger}_\textbf{k} \\
 \end{pmatrix}
{\bf H}_{\bf k}
 \begin{pmatrix}
 \hat a_\textbf{k}  \\
 \hat b_\textbf{k}  \\
 \hat c_\textbf{k}  \\
 \end{pmatrix},
\end{eqnarray}
where
\begin{eqnarray}
{\bf H}_{\bf k}&=&\begin{pmatrix}
  0 & 2\kappa_y\cos(k_y) & 2\kappa_x\cos(k_x) \\
  2\kappa_y\cos(k_y) & 0 & 0 \\
  2\kappa_x\cos(k_x)  & 0 & 0 \\
 \end{pmatrix}.
\end{eqnarray}
The energy spectrum consists of three bands, two dispersive and one flat band. The eigenvectors have the form
\begin{equation}
|\psi^{\pm,0}(\textbf{k})\rangle=( A^{\pm,0}_{\textbf{k}} \hat{a}^{\dagger}_{\textbf{k}}+B^{\pm,0}_{\textbf{k}} \hat{b}^{\dagger}_{\textbf{k}}+C^{\pm,0}_{\textbf{k}} \hat{c}^{\dagger}_{\textbf{k}}) |0 \rangle ,
\notag
\end{equation}
where the superscripts $\pm$ denote the dispersive bands and $0$ denotes the flat band. 
The eigenvectors can be Fourier transformed into real space, giving
\begin{eqnarray}
\label{KEigen}
|\psi^{\pm,0}(\textbf{x})\rangle&=&\frac{1}{\sqrt{N}} \bigg( \sum_{\textbf{R}_a} e^{-i\textbf{k}.\textbf{R}_a}A^{\pm,0}_{\textbf{k}} \hat{a}^{\dagger}_{\textbf{R$_a$}}\\
&&+\sum_{\textbf{R$_b$}} e^{-i\textbf{k}.\textbf{R$_b$}}B^{\pm,0}_{\textbf{k}} \hat{b}^{\dagger}_{\textbf{R$_b$}}\nonumber\\
&&+ \sum_{\textbf{R$_c$}} e^{-i\textbf{k}.\textbf{R$_c$}}C^{\pm,0}_{\textbf{k}} \hat{c}^{\dagger}_{\textbf{R$_c$}}\bigg) |0 \rangle ,
\nonumber
\end{eqnarray}
where $N$ is the number of unit cells and $\textbf{R}_a$, $\textbf{R}_b$ and $\textbf{R}_c$ are the lattice positions of the different A, B and C sites. These eigenvectors form a complete orthonormal basis for real space. 
It can be shown that the $A_\textbf{k}^\pm, B_\textbf{k}^\pm$ and $C_\textbf{k}^\pm$ coefficients for the two dispersive bands are
\begin{eqnarray}
A^{\pm}_\textbf{k}&=& {\pm \frac{\cos(k_x)}{\sqrt{1+\cos(2k_x)}}},
\notag
\\
B^{\pm}_\textbf{k}&=& { \frac{\kappa_y \cos(k_y)\sec(k_x)}{\sqrt{(\kappa_x^2+\kappa_y^2+\kappa_x^2 \cos(2k_x)+\kappa_y^2 \cos(2k_y))\sec(k_x)^2}}} ,
\notag
\\
&\equiv& {\frac{b_k}{f(k_x,k_y)}}
\notag,\\
C^{\pm}_\textbf{k}&=& {\frac{\kappa_x}{\sqrt{(\kappa_x^2+\kappa_y^2+\kappa_x^2 \cos(2k_x)+\kappa_y^2 \cos(2k_y))\sec(k_x)^2}}},
\notag
\\
&\equiv& {\frac{c_k}{f(k_x,k_y)}}.
\notag
\end{eqnarray}
The two dispersive bands differ only in their $A_\textbf{k}$ coefficients.
We denote the non-diffracting input state in the experiment by $|\phi_0 \rangle$, with
\begin{eqnarray}
\psi^{\pm,0}_\textbf{k}&=& \langle \psi^{\pm,0}(\textbf{k})|\phi_0 \rangle.
\notag
\end{eqnarray}
The non-diffracting input state in the experiment has +1 on two C sites and -1 on two B sites. We will now show that this state has zero overlap with the dispersive-band states. Setting the origin on any A site gives, up to an unimportant global shift in phase,
\begin{eqnarray}
\psi^{\pm,0}_\textbf{k}&=&\frac{1}{\sqrt{N}}  (e^{-ik_x}C^{\pm,0}_k+e^{-ik_x-2ik_y}C^{\pm,0}_k-\nonumber\\&&
e^{-ik_y}B^{\pm,0}_k-e^{-2ik_x-ik_y}B^{\pm,0}_k).
\label{Overlap}
\end{eqnarray}
Equation ($\ref{Overlap}$) applies to coefficients for all three bands. For the coefficients for states in the dispersive bands, we obtain
\begin{eqnarray}
 \psi^{\pm}_\textbf{k}&=&\frac{1}{\sqrt{N} f(k_x,k_y)} (e^{-ik_x}c_k(1+e^{-2ik_y})-\nonumber\\&& 
 b_k e^{-ik_y}(1+e^{-2ik_x}))
\notag
\end{eqnarray}
where
\begin{equation}
b_k={\frac{2\kappa_y \cos(k_y)}{e^{ik_x}(1+e^{-2ik_x})}}.
\notag
\end{equation}
Therefore 
\begin{eqnarray}
\psi^{\pm}_\textbf{k} &=& \frac{1}{\sqrt{N} f(k_x,k_y)} \left[e^{-ik_x}c_k(1+e^{-2ik_y})-\right.\nonumber\\
&&\left. b_k e^{-ik_y}(1+e^{-2ik_x})\right]
\notag\\
&=&\frac{1}{\sqrt{N} f(k_x,k_y)} \left[e^{-ik_x-ik_y}\kappa_x(e^{ik_y}+e^{-ik_y})-\right.\nonumber\\
&&\left. 2 \kappa_y \cos(k_y) e^{-ik_y-ik_x}(1+e^{-2ik_x})\right]
\notag
\\
 &=& \frac{2e^{-ik_x-ik_y}\cos(k_y)}{\sqrt{N} f(k_x,k_y)} (\kappa_x -\kappa_y ).
\notag
\end{eqnarray}
When $\kappa_x$=$\kappa_y$ there is therefore no overlap between the non-diffracting input state in the experiment and states in either of the dispersive bands. Consequently, the non-diffracting state must be composed of only flat-band eigenvectors. 


\section{2. Disorder and the flat band}

In this section we show that in the case of a Lieb Hamiltonian $H_1$ with random couplings, that is, off-diagonal disorder, there are $N$ eigenvectors which satisfy the equation $H_1|\eta_i \rangle=0$. Hence these $N$ eigenvectors form a flat band. $N$ is the number of unit cells. Off-diagonal disorder is known to be present in the experiment.

\textbf{Definitions:}
 Let $|\psi^0(\textbf{k})\rangle$ and $|\psi^{\pm}(\textbf{k})\rangle$ be eigenvectors of the disorder-free Hamiltonian $\hat H_0$. The superscript refers to the band. Let $\hat H_1$ be the disordered Hamiltonian. In a finite but periodic lattice, there are only a discrete number of allowed quasimomenta. These are labelled $\textbf{k}_i$, and this label is often summed over. 
\subsection{Outline}
The mechanism at the heart of the derivation is that a zero-energy eigenstate must satisfy
\begin{eqnarray}
\hat H_1|\eta_i\rangle &=&\hat H_1\sum_\textbf{k}\left[ c^0(\textbf{k}) |\psi^0(\textbf{k})\rangle
+ c^+(\textbf{k}) |\psi^+(\textbf k)\rangle \right. \nonumber\\
&&+\left. c^-(\textbf{k}) |\psi^-(\textbf{k})\rangle \right] =0. \label{ZEEigen}
\end{eqnarray}
The key steps  are
\begin{enumerate}
\item We prove that particle-hole symmetry holds also for the disordered lattice. One can then show that $\hat H_1 |\psi^0(k)\rangle$ can only contain population on A sites.

 \item The constraint in 1. means that $\sum_\textbf{k}[ c^+(\textbf{k}) |\psi^+(\textbf{k})\rangle+ 
 c^-(\textbf{k}) |\psi^-(\textbf{k})\rangle]$ can only contain population in A sites if $|\eta_i\rangle$ is to satisfy equation $\eqref{ZEEigen}$.
 
  \item  The constraint in 2. implies that
  \begin{eqnarray}
\sum_\textbf{k}[ c^+(\textbf{k}) |\psi^+(\textbf{k})\rangle+ c^-(\textbf{k}) |\psi^-(\textbf{k})\rangle]\\
=  \sum_\textbf{k} c(\textbf{k})[ |\psi^+(\textbf{k})\rangle+ |\psi^-(\textbf{k})\rangle].\nonumber
  \end{eqnarray}
  

  \item The vectors $\hat H_1[ |\psi^+(\textbf{k})\rangle+ |\psi^-(\textbf{k})\rangle]$ form a basis for the A sites.
  
\item Points 1.-4. show that there exist zero-energy eigenfunctions of the form $|\eta_i \rangle=|\psi^0(\textbf{k}_i)\rangle+\sum_\textbf{k} c_i(\textbf{k})(|\psi^+(\textbf{k})\rangle+|\psi^-(\textbf{k}) \rangle) $.
\end{enumerate}
 More specifically, point 5 can be seen in the following way.
It follows from point 1. that acting with $\hat H_1$ on the first term on the rhs of $|\eta_i \rangle$,  $\sum_\textbf{k} c^0(\textbf{k}) |\psi^0(\textbf{k})\rangle$, gives a vector that only has A site population.  By points 2 and 3, acting with $\hat H_1$ on the remaining part of $|\eta_i\rangle$ 
gives $\sum_\textbf{k} c(\textbf{k}) [\hat H_1( |\psi^+(\textbf{k}_i)\rangle+ |\psi^-(\textbf{k}_i)\rangle)]$. Item 4 says that $\hat H_1[ |\psi^+(\textbf{k})\rangle+ |\psi^-(\textbf{k})\rangle]$ form a basis for the A sites. Therefore, any state with A site population can be written in the form $\sum_\textbf{k} c(\textbf{k}) [\hat H_1( |\psi^+(\textbf{k}_i)\rangle+ |\psi^-(\textbf{k}_i)\rangle)]$.
In particular, by a suitable choice of $c(\textbf{k})$, the A site population coming from $\hat H_1|\psi^0(\textbf{k})\rangle$ can be cancelled, giving $\hat H_1|\eta_i \rangle=0$ as desired. 

\subsection{Disorder and particle-hole symmetry}

\textbf{Claim.}
In the case of  off-diagonal disorder, particle-hole symmetry is still maintained.

\textbf{Proof.} 
The disordered Hamiltonian can be written
\begin{multline}
\hat H_1=\sum_{nm}[(J^+_{x,(n,m)}) \hat c^{\dagger}_{n,m}\hat a_{nm}+(J^-_{x,(n,m)})\hat c^{\dagger}_{n-1,m}\hat a_{n,m}\\
+(J^+_{y,(n,m)})\hat b^{\dagger}_{n,m}\hat a_{n,m}+(J^-_{y,(n,m)})\hat b^{\dagger}_{n,m-1}\hat a_{n,m}]+\text{h.c.}
\notag
\end{multline}
where $J^+$ refers to 
hopping within the same unit cell whilst $J^-$ refers to 
hopping to a different unit cell. The subscripts are needed on the hoppings due to the disorder.
The Schr\"odinger equation for the system is given by
\begin{multline}
\notag
-i \frac{\partial |\xi \rangle}{\partial t}=\hat{H}_1 |\xi \rangle \Leftrightarrow \\
-i \sum_{n,m}(\dot{a}_{n,m}\hat a^{\dagger}_{n,m}+\dot{b}_{n,m} \hat b^{\dagger}_{n,m}+\dot{c}_{n,m} c^{\dagger}_{n,m}) |0 \rangle=\hat{H}_1 |\xi \rangle
\notag
\end{multline}
The rhs of the above equation is
\begin{eqnarray}
\notag
\hat H_1 |\xi \rangle&=&\hat H_1\sum_{n,m}(a_{n,m}\hat a^{\dagger}_{n,m}+b_{n,m} \hat b^{\dagger}_{n,m}+c_{n,m} c^{\dagger}_{n,m}) |0 \rangle\notag\\
&=&\sum_{n,m,p,q} [(J^+_{x,(p,q)}) \hat c^{\dagger}_{p,q}\hat a_{pq}+(J^-_{x,(p,q)})\hat c^{\dagger}_{p-1,q}\hat a_{p,q}\notag\\
&&+(J^+_{y,(p,q)})\hat b^{\dagger}_{p,q}\hat a_{p,q}+(J^-_{y,(p,q)})\hat b^{\dagger}_{p,q-1}\hat a_{p,q}\notag \\
&&+(J^+_{x,(p,q)})\hat a^{\dagger}_{pq} \hat c_{p,q}+(J^-_{x,(p,q)})\hat a^{\dagger}_{p,q}\hat c_{p-1,q}\notag\\
&&+(J^+_{y,(p,q)})\hat a^{\dagger}_{p,q}\hat b_{p,q}+(J^-_{y,(p,q)})\hat a^{\dagger}_{p,q}\hat b_{p,q-1}]\notag\\
&&\times (a_{n,m}\hat a^{\dagger}_{n,m}+b_{n,m} \hat b^{\dagger}_{n,m}+c_{n,m} \hat c^{\dagger}_{n,m})|0 \rangle \notag \\
&=&\sum_{n,m} [(J^+_{x,(n,m)}) a_{n,m} \hat c^{\dagger}_{n,m}+(J^-_{x,(n,m)})a _{n,m}\hat c^{\dagger}_{n-1,m}\notag\\
&&+(J^+_{y,(n,m)})a_{n,m}\hat b^{\dagger}_{n,m}+(J^-_{y,(n,m)})a_{n,m}\hat b^{\dagger}_{n,m-1} \label{HamState}\\
&&+(J^+_{x,(n,m)})c_{n,m}\hat a^{\dagger}_{n,m} +(J^-_{x,(n+1,m)})c_{n,m}\hat a^{\dagger}_{n+1,m}\notag\\
&&+(J^+_{y,(n,m)})b_{n,m}\hat a^{\dagger}_{n,m}+(J^-_{y,(n,m+1)})b_{n,m}\hat a^{\dagger}_{n,m+1}]|0 \notag \rangle .
\end{eqnarray}
Acting from the left with the different annihilation operators upon equation $\eqref{HamState}$ gives the following equations of motion for the probability amplitudes:
\begin{eqnarray}
-i\dot{a}_{n,m}&=& -J^{+}_{x,(n,m)} c_{n,m}-J^{-}_{x,(n,m)}c_{n-1,m}\notag \\
&&-J^{+}_{y,(n,m)} b_{n,m}-J^{-}_{y,(n,m)} b_{n,m-1} \label{Amotion}\\
-i\dot{b}_{n,m}&=&J^{+}_{y,(n,m)} a_{n,m}+J^{-}_{y,(n,m+1)}a_{n,m+1} \label{Bmotion}\\
-i\dot{c}_{n,m}&=& J^{+}_{x,(n,m)} a_{n,m}+J^{-}_{x,(n+1,m)} a_{n+1,m}.\label{Cmotion}
\end{eqnarray}
For a state to be an eigenvector of $H_1$ its probability amplitudes must satisfy the equations
\begin{eqnarray}
E a_{n,m}&=&J^{+}_{x,(n,m)}c_{n,m}+J^{-}_{x,(n,m)}c_{n-1,m}+\nonumber\\
&& J^{+}_{y,(n,m)}b_{n,m}+J^{-}_{y,(n,m)}b_{n,m-1} \label{AEigen}\\
\label{BEigen}
E b_{n,m}&=&J^{+}_{y,(n,m)}a_{n,m}+J^{-}_{y,(n,m+1)}a_{n,m+1}\\
\label{CEigen}
E c_{n,m}&=& J^{+}_{x,(n,m)}a_{n,m}+J^{-}_{x,(n+1,m)}a_{n+1,m} .
\end{eqnarray}
Consider a new vector with the elements $a_{n,m}'$,$b_{n,m}'$ and $c_{n,m}'$, where $b_{n,m}'$ and $c_{n,m}'$ are the same as $b_{n,m}$ and $c_{n,m}$ for the eigenvector but  $a_{n,m}'=-a_{n,m}$. Equation $\eqref{Amotion}$ gives
\begin{eqnarray}
-i\dot{a'}_{n,m}&=& J^{+}_{x,(n,m)}c'_{n,m}+J^{-}_{x,(n,m)}c'_{n-1,m} +\nonumber\\&&
J^{+}_{y,(n,m)}b'_{n,m}+J^{-}_{y,(n,m)}b'_{n,m-1}\nonumber\\
&=&J^{+}_{x,(n,m)}c_{n,m}+J^{-}_{x,(n,m)}c_{n-1,m} +\nonumber\\&& 
J^{+}_{y,(n,m)}b_{n,m}+J^{-}_{y,(n,m)}b_{n,m-1} \nonumber\\
&=& Ea_{n,m}\nonumber\\
&=&-E  a'_{n,m}.
\notag
\end{eqnarray}
 Equation $\eqref{Cmotion}$ gives
\begin{eqnarray}
\notag
-i\dot{c'}_{n,m}&=&J^{+}_{x,(n,m)}a'_{n,m}+J^{-}_{x,(n+1,m)}a'_{n+1,m}\\
\notag
&=& ( -J^{+}_{x,(n,m)}a_{n,m}-J^{-}_{x,(n+1,m)}a_{n+1,m})\\
\notag
&=& -( J^{+}_{x,(n,m)}a_{n,m}+J^{-}_{x,(n+1,m)}a_{n+1,m})\\
\notag
&=& -(E c_{n,m})\\
\notag
&=&(-E) c_{n,m}\\
\notag
&=&(-E) c_{n,m}'.
\end{eqnarray}
A similar equation as for $c_{n,m}$ holds for $b_{n,m}$. 
Therefore, the new vector with elements $a_{n,m}'$,$b_{n,m}'$ and $c_{n,m}'$ is also an eigenvector with energy $-E$.
For an eigenvector of the disordered Hamiltonian with energy $E$ there exists another eigenvector with energy $-E$, which is particle-hole symmetry. This new eigenvector is obtained from the original one by sending $a_{n,m}\rightarrow -a_{n,m}$ for all the the A sites. The operator $\hat U$ that performs this operation therefore leaves B and C sites unchanged. This operator anticommutes with the Hamiltonian since
\begin{eqnarray}
\hat U(\hat H_1 |\psi \rangle)&=& E \hat U  |\psi \rangle,
\label{PHSym1}\\
\hat H_1 (\hat U|\psi \rangle)&=& -E \hat U|\psi \rangle.
\label{PHSym2}
\end{eqnarray}
In equation $\eqref{PHSym2}$, we used particle-hole symmetry. Adding equations $\eqref{PHSym1}$ and $\eqref{PHSym2}$ gives
\begin{equation}
(\hat U\hat H_1+\hat H_1\hat U)|\psi \rangle=0.
\notag
\end{equation}

\subsection{Probability on  A-sites in $\hat H_1|\psi^0(\textbf{k})\rangle$}
In this section it is shown that $\hat H_1|\psi^0(\textbf{k})\rangle$  only has occupation on A-sites.

\textbf{Claim.}
$\hat H_1|\psi^0(\textbf{k})\rangle$ only has probability amplitude on A-sites.

\textbf{Proof.}
Consider the effect $\hat H_1$ has on a flat-band eigenvector, $|\psi^0(\textbf{k})\rangle$:
\begin{eqnarray}
|\phi \rangle=&\hat H_1 |\psi^0(\textbf{k})\rangle \nonumber\\
=& \hat H_1 \hat U|\psi^0(\textbf{k})\rangle \\
\label{FlatU}
=&-\hat U\hat H_1 |\psi^0(\textbf{k})\rangle \nonumber\\
=&-\hat U|\phi \rangle.
\label{PHFlat}
\end{eqnarray}
 Here we used the fact that $\hat U$ only effects A-sites, and that $|\psi^0 (\textbf{k})\rangle$ has zero probability amplitude for all A sites. 
The operator $\hat U$ leaves B and C sites unchanged. Therefore, for an arbitrary B-site component, $\eqref{PHFlat}$ gives $b_{n,m}=-b_{n,m}\rightarrow b_{n,m}=0$ and similarly for the C-site components. This means that $|\phi \rangle$ can only have A site population.

\textbf{Consequences.}
Any eigenvector of the disordered Hamiltonian can be written in terms of the eigenvectors for the disorder-free Hamiltonian as
\begin{multline}
|\eta^0 \rangle= \sum_k c^0(\textbf{k}) |\psi^0(\textbf{k})\rangle+ \sum_\textbf{k} c^+(\textbf{k}) |\psi^+(\textbf{k})\rangle \\
+\sum_\textbf{k} c^-(\textbf{k}) |\psi^-(\textbf{k})\rangle.
\notag
\end{multline}
For this eigenvector to be a zero-energy eigenvector of the disordered Hamiltonian, it is required that
\begin{multline}
0=\hat H_1\left[\sum_\textbf{k} c^0(\textbf{k}) |\psi^0(\textbf{k})\rangle+ \sum_k c^+(\textbf{k}) |\psi^+(\textbf{k}) \right.\rangle\\
\left. + \sum_\textbf{k} c^-(\textbf{k}) |\psi^-(k)\rangle\right].
\label{Eigenval}
\end{multline}
Therefore, since $\hat H_1 |\psi^0(\textbf{k})\rangle$ can only have A-site population, for equation $\eqref{Eigenval}$ to hold, it is  required that $\hat H_1[\sum_\textbf{k} c^+(\textbf{k}) |\psi^+(\textbf{k})\rangle+ \sum_\textbf{k} c^-(\textbf{k}) |\psi^-(\textbf{k})\rangle]$ should only have A-site population. 
\\
\subsection{Form of dispersive band component}
In the previous section it was shown that for $|\eta^0 \rangle$,  $\hat H_1\sum_\textbf{k}[c^+(\textbf{k}) |\psi^+(\textbf{k})\rangle+  c^-(\textbf{k}) |\psi^-(\textbf{k})\rangle]$ should only have A-site population.  In this section we show that this forces the  $\sum_\textbf{k} c^+(\textbf{k}) |\psi^+(\textbf{k})\rangle+ c^-(\textbf{k}) |\psi^-(\textbf{k})\rangle$ component of $|\eta^0 \rangle$ to be equal to $\sum_\textbf{k} c(\textbf{k})[ |\psi^+(\textbf{k})\rangle+ |\psi^-(\textbf{k})\rangle]$.
The proof of this fact relies upon the following claim.

\textbf{Claim.}
$[ c^+(k) |\psi^+(k)\rangle+  c^-(k) |\psi^-(k)\rangle]$ having no A-site population is  a solution to the requirement that $ \hat H_1[ c^+(k) |\psi^+(k)\rangle+ c^-(k) |\psi^-(k)\rangle]$ only have A-site population.

\textbf{Proof.}
 $\sum_k[ c^+(k) |\psi^+(k)\rangle+  c^-(k) |\psi^-(k)\rangle]$ has a general form given by  $|\xi \rangle=\sum_{n,m}(a_{n,m}\hat a^{\dagger}_{n,m}+b_{n,m} \hat b^{\dagger}_{n,m}+c_{n,m} \hat c^{\dagger}_{n,m})|0 \rangle$.
It was shown in equation $\eqref{HamState}$ that $H_1$ acting upon this state gives
\begin{eqnarray}
\hat H_1 |\xi \rangle&=
&\sum_{n,m} [(J^+_{x,(n,m)}) \lambda_{n,m} \hat c^{\dagger}_{n,m}+(J^-_{x,(n,m)})\lambda_{n,m} \hat c^{\dagger}_{n-1,m}\notag\\
&&+(J^+_{y,(n,m)})a_{n,m}\hat b^{\dagger}_{n,m}+(J^-_{y,(n,m)})a_{n,m}\hat b^{\dagger}_{n,m-1}\label{ASiteProof}\\
&&+(J^+_{x,(n,m)})c_{n,m}\hat a^{\dagger}_{n,m} +(J^-_{x,(n+1,m)})c_{n,m}\hat a^{\dagger}_{n+1,m}\notag\\
&&+(J^+_{y,(n,m)})b_{n,m}\hat a^{\dagger}_{n,m}+(J^-_{y,(n,m+1)})b_{n,m}\hat a^{\dagger}_{n,m+1}]|0 \rangle . \notag
\end{eqnarray}
 Lines 1 and 2 of Eq. $\eqref{ASiteProof}$ show that the probability amplitude on the B and C sites of the vector $\hat H_1 |\xi \rangle $ is proportional to $a_{n,m}$, which is $|\xi \rangle$'s A-site probability amplitude. Therefore, if $\sum_\textbf{k}[ c^+(\textbf{k}) |\psi^+(\textbf{k})\rangle+  c^-(\textbf{k}) |\psi^-(\textbf{k})\rangle]$ contains no A-site probability then this satisfies the requirement that $\hat H_1 (\sum_\textbf{k}[ c^+(\textbf{k}) |\psi^+(\textbf{k})\rangle+  c^-(\textbf{k}) |\psi^-(\textbf{k})\rangle]) $ contain only A-site probability.

We are now in a position to show $c^+(\textbf{k})=c^-(\textbf{k})$.

\textbf{Claim.}
For $\sum_\textbf{k}[ c^+(\textbf{k}) |\psi^+(\textbf{k})\rangle+  c^-(\textbf{k}) |\psi^-(\textbf{k})\rangle]$ to have no A-site population it is required that $c^+(\textbf{k})=c^-(\textbf{k})$.

\textbf{Proof.}
 In the disorder-free case the dispersive bands have eigenvectors of the form
\begin{multline}
\notag
|\psi^{\pm} (\textbf{k})\rangle= \sum_{n,m} (\pm A_\textbf{k} e^{i\textbf{k}.\textbf{R}^A_{nm}}\hat a^{\dagger}_{n,m}+B_\textbf{k} e^{i\textbf{k}.\textbf{R}^B_{nm}}\hat b^{\dagger}_{n,m}\\
+C_k e^{i\textbf{k}.\textbf{R}^C_{nm}}\hat c^{\dagger}_{n,m})|0 \rangle.
\end{multline}
It is required that there is no A-site population. Therefore, for arbitrary $p$ and $q$ it must hold that 
\begin{eqnarray}
0&=& \langle 0| \hat a_{pq}\left[\sum_\textbf{k}(c^+(\textbf{k})|\psi^+(\textbf{k})\rangle+c^-(\textbf{k})|\psi^-(\textbf{k})\rangle\right]\nonumber\\
&=& \sum_k\left[c^+(\textbf{k})A_\textbf{k} e^{i\textbf{k}.\textbf{R}^A_{pq}}-c^-(\textbf{k})A_\textbf{k} e^{i\textbf{k}.\textbf{R}^A_{pq}}\right]\nonumber\\
&=&\sum_\textbf{k}A_\textbf{k} e^{i\textbf{k}.\textbf{R}^A_{pq}}\left[c^+(\textbf{k})-c^-(\textbf{k})\right].
\notag
\end{eqnarray}
This has to hold for all $\textbf{R}^A$ and so $c^+(\textbf{k})=c^{-}(\textbf{k})$. 
The consequence of $c^+(\textbf{k})=c^{-}(\textbf{k})$ is that the expression for the zero-energy eigenstate becomes
\begin{equation}
|\eta^0 \rangle=\sum_\textbf{k} c^0(\textbf{k}) |\psi^0(\textbf{k})\rangle+ \sum_\textbf{k} c(\textbf{k}) \left[|\psi^+(\textbf{k})\rangle+ |\psi^-(\textbf{k})\rangle\right].
\notag
\end{equation}

\subsection{Basis for A sites}
In the previous section it was shown that  if a zero-enegy state exists it must have the form
\begin{equation}
\notag
|\eta^0(\textbf{k}) \rangle=\sum_\textbf{k} c^0(\textbf{k}) |\psi^0(\textbf{k})\rangle+ \sum_\textbf{k} c(\textbf{k}) \left[|\psi^+(\textbf{k})\rangle+ |\psi^-(\textbf{k})\rangle\right].
\end{equation}
Consider acting with $\hat H_1$ upon this state. It has been shown that $\hat H_1 |\psi^0(\textbf{k})\rangle$ produces a vector with only A-site population, and in the previous section it was shown that $\hat H_1[|\psi^+(\textbf{k})\rangle+ |\psi^-(\textbf{k})\rangle]$ also has only A-site population. This suggests that it may be possible through some appropriate choice of $c(\textbf{k})$ to cancel the terms coming from $\hat H_1 |\psi^0(\textbf{k})\rangle$. If this were the case then $\hat H_1|\eta^0(\textbf{k})\rangle$ would be zero and we would have a zero-energy eigenstate.

To show that this cancellation is possible we must show that the set of vectors $\hat H_1[|\psi^+(\textbf{k}_i)\rangle+ |\psi^-(\textbf{k})\rangle]$, which we now call $|\mu^\textbf{k} \rangle$,  are a basis for the A sites. 
 In the disorder-free case there are as many values of $k$ as there are unit cells, $N$, and so there are $N$ different $|\mu^\textbf{k} \rangle$ states. 
There are $N$ A sites in the lattice, and so if the $|\mu^{\textbf{k}_i} \rangle$ are linearly independent then $|\mu^\textbf{k}\rangle$ will form a basis for the A sites. 

\textbf{Claim.}
The $|\mu^{\textbf{k}_i} \rangle$ are linearly independent.

\textbf{Proof.}
To test for linear independence it must be shown that the only solution to $\sum_i \alpha_{\textbf{k}_i} |\mu^{\textbf{k}_i} \rangle=0$ is $\alpha_{\textbf{k}_i}=0$  $\forall i$.
Therefore the equation under consideration is 
\begin{equation}
\notag
\sum_i \alpha_{\textbf{k}_i} \hat H_1\left[|\psi^+(\textbf{k})\rangle+  |\psi^-(\textbf{k})\rangle\right]=0.
\end{equation}
 If a solution would exist which does not have $\alpha_{\textbf{k}_i} \neq 0$ $\forall i$  then this solution, $|\phi \rangle=\sum_i \alpha_{\textbf{k}_i}(|\psi^+(\textbf{k}_i)\rangle+  |\psi^-(\textbf{k}_i)\rangle)$,  satisfies $\hat H_1 |\phi \rangle=0$. Therefore $|\phi \rangle$  would represent a zero-energy eigenstate.
 If it can be shown that $\sum_i \alpha_{\textbf{k}_i}(|\psi^+(\textbf{k})\rangle+  |\psi^-(\textbf{k})\rangle)$ cannot be a zero-energy eigenvector, then this would prove that that the $ |\mu^{\textbf{k}_i} \rangle$ are linearly independent and therefore a basis for the A sites. As has been discussed previously, $ |\mu^{\textbf{k}_i} \rangle$ only has A-site population. For $ |\mu^{\textbf{k}_i} \rangle$ to be a zero-energy eigenvector we obtain from equations $\eqref{CEigen}$ and $\eqref{BEigen}$ that
\begin{eqnarray}
\label{ASiteEigenx}
0=& -J^{+}_{x,(n,m)}a_{n,m}-J^{-}_{x,(n+1,m)}a_{n+1,m},\\
0=&-J^{+}_{y,(n,m)}a_{n,m}-J^{-}_{y,(n,m+1)}a_{n,m+1}.
\label{ASiteEigeny}
\end{eqnarray}
Let us assume that these equations hold and let the probability amplitude on the $(n,m)^{th}$ A-site be $x$. Then Eqn. \eqref{ASiteEigeny} gives
\begin{eqnarray}
a_{n,m+1}= \frac{-J_{y,(n,m)}^+}{J_{y,(n,m+1)}^-}x.
\notag
\end{eqnarray}
This means that $a_{n,m+1}$ is now determined by $x$. In turn, $a_{n,m+1}$ determines $a_{n+1,m+1}$ via Eqn. \eqref{ASiteEigenx},
\begin{eqnarray}
\notag
a_{n+1,m+1}=&  \frac{-J_{x,(n,m+1)}^+}{J_{x,(n+1,m+1)}^-}a_{n,m+1}\\
=& \frac{-J_{x,(n,m+1)}^+}{J_{x,(n+1,m+1)}^-}\cdot\frac{-J_{y,(n,m)}^+}{J_{y,(n,m+1)}^-}x.
\label{Path2}
\end{eqnarray}
The probability amplitude $x$ also determines $a_{n+1,m}$ via equation \eqref{ASiteEigenx},
\begin{eqnarray}
\notag
a_{n+1,m}= \frac{-J_{x,(n,m)}^+}{J_{x,(n+1,m)}^-}x
\end{eqnarray}
and $a_{n+1,m}$ determines $a_{n+1,m+1}$ via equation $\eqref{ASiteEigeny}$,
\begin{eqnarray}
\notag
a_{n+1,m+1}=& \frac{-J_{y,(n+1,m)}^+}{J_{x,(n+1,m+1)}^-}a_{n+1,m},\\
=& \frac{-J_{y,(n+1,m)}^+}{J_{x,(n+1,m+1)}^-}\cdot\frac{-J_{x,(n,m)}^+}{J_{x,(n+1,m)}^-}x.
\label{Path1}
\end{eqnarray}
Therefore, this gives us two equations for what $a_{n+1,m+1}$ must be if $a_{n,m}=x$.
Equating \eqref{Path1} and \eqref{Path2} gives
\begin{equation}
 \frac{J_{y,(n+1,m)}^+}{J_{y,(n+1,m+1)}^-}\cdot\frac{J_{y,(n,m+1)}^-}{J_{y,(n,m)}^+}=\frac{J_{x,(n,m+1)}^+}{J_{x,(n+1,m+1)}^-}\cdot\frac{J_{x,(n+1,m)}^-}{J_{x,(n,m)}^+}.
\label{Bonds}
\end{equation}
Equation \eqref{Bonds} says that for the solution $a_{n,m}=x$ to be consistent with the eigenvector equations, it is required that the expression of the bonds in the $x$ direction in the lhs and the expression of the bonds in the $y$ direction in the rhs be equal. For a randomly disordered Hamiltonian, this will in general not hold. If equation \eqref{Bonds} does not hold then the only solution that avoids a contradiction is $x=0$. This means that for a disordered Hamiltonian, a state with only A-site population cannot be a zero-energy state. Therefore, the $ |\mu^{\textbf{k}_i} \rangle$  are linearly independent. 

\subsection{Zero-energy eigenstates}
The consquence of the  $ |\mu^{\textbf{k}_i} \rangle$ being linearly independent is that the $|\mu^\textbf{k}_i \rangle$ form a basis for the A sites. This is crucial as it allows the creation of zero-energy eigenstates of the disordered lattice. A zero-energy eigenstate has the form
\begin{equation}
|\eta^0 \rangle=|\psi^0(\textbf{k})\rangle+\sum_\textbf{k} c(\textbf{k})(|\psi^+(\textbf{k})\rangle+|\psi^-(\textbf{k}) \rangle).
\label{EigenForm}
\end{equation}
A zero-energy eigenvector must satisfy $\hat H_1|\eta^0\rangle=0$, and this holds for the state of equation \eqref{EigenForm} as shown now. Acting with $\hat H_1$ on the first term on the rhs of \eqref{EigenForm} gives a vector that only has A site population. Acting with $\hat H_1$ on the second term gives $\sum_\textbf{k} c(\textbf{k}) |\mu^k \rangle$ where $|\mu^\textbf{k} \rangle$ form a basis for the A sites. Therefore, by a suitable choice of $c(\textbf{k})$, the A site population coming from $\hat H_1|\psi^0(\textbf{k})\rangle$ can be cancelled giving $\hat H_1|\eta^0 \rangle=0$ as desired.

\subsection{Flat band}
 A whole family of such zero-energy eigenvectors can be created by changing the value of $\textbf{k}$ that is used for $|\psi^0(\textbf{k})\rangle$. A general member of this family has the form
\begin{equation}
|\eta^0_i\rangle=|\psi^0(\textbf{k}_i)\rangle+\sum_\textbf{k} c_i(\textbf{k})(|\psi^+(\textbf{k})\rangle+|\psi^-(\textbf{k}) \rangle).
\notag
\end{equation}

\textbf{Claim.}
The $|\eta^0_i\rangle$ are linearly independent.

\textbf{Proof.}
\begin{eqnarray}
\notag
0&=& \sum_i \alpha_i |\eta^0_i\rangle\\
&=& \sum_i\alpha_i[|\psi^0(\textbf{k}_i)\rangle+\sum_k c_i(\textbf{k})(|\psi^+(\textbf{k})\rangle+|\psi^-(\textbf{k}) \rangle)].\nonumber
\end{eqnarray}
The term in the brackets is a  linear combination of disorder-free eigenstates, and therefore it cannot be zero as this would mean that the disorder-free eigenvectors were not linearly independent. The other possibility is that the brackets corresponding to two different $\textbf{k}$ values cancel. Therefore assume that there is a solution with $\alpha_i \neq 0$ for some $i$. Let one of the non-zero $\alpha_i$ be $\alpha_0$. Then
\begin{multline}
\notag
0= \alpha_0 |\psi^0(\textbf{k}_0) \rangle+\alpha_0 \sum_\textbf{k} c_0(\textbf{k})(|\psi^+(\textbf{k})\rangle+|\psi^-(\textbf{k}) \rangle)\\
+\sum_{i,i\neq 0}\alpha_i[|\psi^0(\textbf{k}_i)\rangle+\sum_\textbf{k} c_i(\textbf{k})(|\psi^+(\textbf{k})\rangle+|\psi^-(\textbf{k}) \rangle)]
\notag
\end{multline}
and
\begin{multline}
\notag
|\psi^0(\textbf{k}_0) \rangle= \beta_0 \sum_\textbf{k} c_0(\textbf{k})(|\psi^+(\textbf{k})\rangle+|\psi^-(\textbf{k}) \rangle)\\
+\sum_{i,i\neq 0}\beta_i[|\psi^0(\textbf{k}_i)\rangle+\sum_\textbf{k} c_i(\textbf{k})(|\psi^+(\textbf{k})\rangle+|\psi^-(\textbf{k}) \rangle)].
\notag
\end{multline}
This expression violates the linear independence of the disorder-free eigenvectors, which are a basis. Therefore, the only solution is $\alpha_i =0$ $\forall i$, and therefore the $|\eta_i \rangle$ are linearly independent. These linearly independent eigenvectors when combined with other linearly independent non-zero energy state can be converted into a orthonormal basis via the Gram-Schmidt process. This procedure does not change the eigenvalues, and so the disordered Hamiltonian always possesses $N$ zero-energy orthonormal eigenvectors,  where $N$ is the number of unit cells. Hence the flat band persists even in the presence of disorder, which was what we wanted to show.

\subsection{Comment on diagonal disorder}

Finally we would like to note that in the case of diagonal disorder it can be seen numerically that the flat band is destroyed by this type of disorder. By numerically calculating the eigenvalues of the disordered lattice, and ensemble averaging, a distribution function for the energy spacing between adjacent eigenvalues was obtained for the eigenvalues close to zero. This distribution function goes to zero as the spacing goes to zero. Therefore there is only a small probability of degenerate eigenvalues and hence the flat band is broken. This repulsion of neighbouring eigenvalues is a well known phenomena in random matrix theory and is described by the Wigner-Dyson distribution.    Furthermore, numerically, it has been observed that the experimental flat band state only weakly disperses in the presence of diagonal disorder. This dispersive behaviour is due to the breaking of the flat band. The breaking of this flat band can be directly related to the diagonal disorder using the Bauer-Fike theorem [F.L. Bauer and C.F. Fike, Numer. Math. 2, 137 (1960)]. The theorem states that the eigenvalues of the perturbed matrix, i.e. the Hamiltonian with disorder, cannot differ from the eigenvalues of the disorder free Hamiltonian by more than the largest eigenvalue of the perturbing Hamiltonian. As the perturbation matrix is diagonal the eigenvalues are easily obtained. Therefore, for weak disorder the flat band is almost maintained. 
    In the very long time limit the experimental state would disperse throughout the lattice for diagonal disorder. The same is not true for off-diagonal disorder. In this case the flat band persists, as shown above, and therefore the projection of the original state onto this flat band will remain localised.


\begin{thebibliography}{33}
\expandafter\ifx\csname natexlab\endcsname\relax\def\natexlab#1{#1}\fi
\expandafter\ifx\csname bibnamefont\endcsname\relax
  \def\bibnamefont#1{#1}\fi
\expandafter\ifx\csname bibfnamefont\endcsname\relax
  \def\bibfnamefont#1{#1}\fi
\expandafter\ifx\csname citenamefont\endcsname\relax
  \def\citenamefont#1{#1}\fi
\expandafter\ifx\csname url\endcsname\relax
  \def\url#1{\texttt{#1}}\fi
\expandafter\ifx\csname urlprefix\endcsname\relax\def\urlprefix{URL }\fi
\providecommand{\bibinfo}[2]{#2}
\providecommand{\eprint}[2][]{\url{#2}}

\bibitem[{\citenamefont{Billy et~al.}(2008)\citenamefont{Billy, Josse, Zuo,
  Bernard, Hambrecht, Lugan, Cl{\'e}ment, Sanchez-Palencia, Bouyer, and
  Aspect}}]{Billy:2008kd}
\bibinfo{author}{\bibfnamefont{J.}~\bibnamefont{Billy}},
  \bibinfo{author}{\bibfnamefont{V.}~\bibnamefont{Josse}},
  \bibinfo{author}{\bibfnamefont{Z.}~\bibnamefont{Zuo}},
  \bibinfo{author}{\bibfnamefont{A.}~\bibnamefont{Bernard}},
  \bibinfo{author}{\bibfnamefont{B.}~\bibnamefont{Hambrecht}},
  \bibinfo{author}{\bibfnamefont{P.}~\bibnamefont{Lugan}},
  \bibinfo{author}{\bibfnamefont{D.}~\bibnamefont{Cl{\'e}ment}},
  \bibinfo{author}{\bibfnamefont{L.}~\bibnamefont{Sanchez-Palencia}},
  \bibinfo{author}{\bibfnamefont{P.}~\bibnamefont{Bouyer}}, \bibnamefont{and}
  \bibinfo{author}{\bibfnamefont{A.}~\bibnamefont{Aspect}},
  \bibinfo{journal}{Nature} \textbf{\bibinfo{volume}{453}},
  \bibinfo{pages}{891} (\bibinfo{year}{2008}).

\bibitem[{\citenamefont{{Schreiber} et~al.}(2015)\citenamefont{{Schreiber},
  {Hodgman}, {Bordia}, {L{\"u}schen}, {Fischer}, {Vosk}, {Altman}, {Schneider},
  and {Bloch}}}]{schreiber2015}
\bibinfo{author}{\bibfnamefont{M.}~\bibnamefont{{Schreiber}}},
  \bibinfo{author}{\bibfnamefont{S.~S.} \bibnamefont{{Hodgman}}},
  \bibinfo{author}{\bibfnamefont{P.}~\bibnamefont{{Bordia}}},
  \bibinfo{author}{\bibfnamefont{H.~P.} \bibnamefont{{L{\"u}schen}}},
  \bibinfo{author}{\bibfnamefont{M.~H.} \bibnamefont{{Fischer}}},
  \bibinfo{author}{\bibfnamefont{R.}~\bibnamefont{{Vosk}}},
  \bibinfo{author}{\bibfnamefont{E.}~\bibnamefont{{Altman}}},
  \bibinfo{author}{\bibfnamefont{U.}~\bibnamefont{{Schneider}}},
  \bibnamefont{and} \bibinfo{author}{\bibfnamefont{I.}~\bibnamefont{{Bloch}}},
  \bibinfo{journal}{ArXiv e-prints}  (\bibinfo{year}{2015}),
  \eprint{1501.05661}.

\bibitem[{\citenamefont{Mielke}(1992)}]{Mielke:1992ct}
\bibinfo{author}{\bibfnamefont{A.}~\bibnamefont{Mielke}},
  \bibinfo{journal}{Journal of Physics A: Mathematical and General}
  \textbf{\bibinfo{volume}{25}}, \bibinfo{pages}{4335} (\bibinfo{year}{1992}).

\bibitem[{\citenamefont{Aoki et~al.}(1996)\citenamefont{Aoki, Ando, and
  Matsumura}}]{Aoki:1996ik}
\bibinfo{author}{\bibfnamefont{H.}~\bibnamefont{Aoki}},
  \bibinfo{author}{\bibfnamefont{M.}~\bibnamefont{Ando}}, \bibnamefont{and}
  \bibinfo{author}{\bibfnamefont{H.}~\bibnamefont{Matsumura}},
  \bibinfo{journal}{Phys. Rev. B} \textbf{\bibinfo{volume}{54}},
  \bibinfo{pages}{1517930} (\bibinfo{year}{1996}).

\bibitem[{\citenamefont{Deng et~al.}(2003)\citenamefont{Deng, Simon, and
  K\"ohler}}]{Deng2003}
\bibinfo{author}{\bibfnamefont{S.}~\bibnamefont{Deng}},
  \bibinfo{author}{\bibfnamefont{A.}~\bibnamefont{Simon}}, \bibnamefont{and}
  \bibinfo{author}{\bibfnamefont{J.}~\bibnamefont{K\"ohler}},
  \bibinfo{journal}{Journal of Solid State Chemistry}
  \textbf{\bibinfo{volume}{176}}, \bibinfo{pages}{412 } (\bibinfo{year}{2003}).

\bibitem[{\citenamefont{Wu et~al.}(2007)\citenamefont{Wu, Bergman, Balents, and
  Das~Sarma}}]{Wu:2007iz}
\bibinfo{author}{\bibfnamefont{C.}~\bibnamefont{Wu}},
  \bibinfo{author}{\bibfnamefont{D.}~\bibnamefont{Bergman}},
  \bibinfo{author}{\bibfnamefont{L.}~\bibnamefont{Balents}}, \bibnamefont{and}
  \bibinfo{author}{\bibfnamefont{S.}~\bibnamefont{Das~Sarma}},
  \bibinfo{journal}{Phys. Rev. Lett.} \textbf{\bibinfo{volume}{99}},
  \bibinfo{pages}{70401} (\bibinfo{year}{2007}).

\bibitem[{\citenamefont{Tasaki}(2008)}]{Tasaki2008}
\bibinfo{author}{\bibfnamefont{H.}~\bibnamefont{Tasaki}},
  \bibinfo{journal}{Eur. Phys. J. B} \textbf{\bibinfo{volume}{64}},
  \bibinfo{pages}{365} (\bibinfo{year}{2008}).

\bibitem[{\citenamefont{Lan et~al.}(2012)\citenamefont{Lan, Goldman, and
  {\"O}hberg}}]{Lan:2012gz}
\bibinfo{author}{\bibfnamefont{Z.}~\bibnamefont{Lan}},
  \bibinfo{author}{\bibfnamefont{N.}~\bibnamefont{Goldman}}, \bibnamefont{and}
  \bibinfo{author}{\bibfnamefont{P.}~\bibnamefont{{\"O}hberg}},
  \bibinfo{journal}{Phys. Rev. B} \textbf{\bibinfo{volume}{85}},
  \bibinfo{pages}{155451} (\bibinfo{year}{2012}).

\bibitem[{\citenamefont{Jacqmin et~al.}(2014)\citenamefont{Jacqmin, Carusotto,
  Sagnes, Abbarchi, Solnyshkov, Malpuech, Galopin, Lema{\^\i}tre, Bloch, and
  Amo}}]{Jacqmin:2014cj}
\bibinfo{author}{\bibfnamefont{T.}~\bibnamefont{Jacqmin}},
  \bibinfo{author}{\bibfnamefont{I.}~\bibnamefont{Carusotto}},
  \bibinfo{author}{\bibfnamefont{I.}~\bibnamefont{Sagnes}},
  \bibinfo{author}{\bibfnamefont{M.}~\bibnamefont{Abbarchi}},
  \bibinfo{author}{\bibfnamefont{D.~D.} \bibnamefont{Solnyshkov}},
  \bibinfo{author}{\bibfnamefont{G.}~\bibnamefont{Malpuech}},
  \bibinfo{author}{\bibfnamefont{E.}~\bibnamefont{Galopin}},
  \bibinfo{author}{\bibfnamefont{A.}~\bibnamefont{Lema{\^\i}tre}},
  \bibinfo{author}{\bibfnamefont{J.}~\bibnamefont{Bloch}}, \bibnamefont{and}
  \bibinfo{author}{\bibfnamefont{A.}~\bibnamefont{Amo}},
  \bibinfo{journal}{Phys. Rev. Lett.} \textbf{\bibinfo{volume}{112}},
  \bibinfo{pages}{116402} (\bibinfo{year}{2014}).

\bibitem[{\citenamefont{Bodyfelt et~al.}(2014)\citenamefont{Bodyfelt, Leykam,
  Danieli, Yu, and Flach}}]{bodyfelt2014}
\bibinfo{author}{\bibfnamefont{J.~D.} \bibnamefont{Bodyfelt}},
  \bibinfo{author}{\bibfnamefont{D.}~\bibnamefont{Leykam}},
  \bibinfo{author}{\bibfnamefont{C.}~\bibnamefont{Danieli}},
  \bibinfo{author}{\bibfnamefont{X.}~\bibnamefont{Yu}}, \bibnamefont{and}
  \bibinfo{author}{\bibfnamefont{S.}~\bibnamefont{Flach}},
  \bibinfo{journal}{Phys. Rev. Lett.} \textbf{\bibinfo{volume}{113}},
  \bibinfo{pages}{236403} (\bibinfo{year}{2014}).

\bibitem[{\citenamefont{Shen et~al.}(2010)\citenamefont{Shen, Shao, Wang, and
  Xing}}]{Shen:2010jz}
\bibinfo{author}{\bibfnamefont{R.}~\bibnamefont{Shen}},
  \bibinfo{author}{\bibfnamefont{L.~B.} \bibnamefont{Shao}},
  \bibinfo{author}{\bibfnamefont{B.}~\bibnamefont{Wang}}, \bibnamefont{and}
  \bibinfo{author}{\bibfnamefont{D.~Y.} \bibnamefont{Xing}},
  \bibinfo{journal}{Phys. Rev. B} \textbf{\bibinfo{volume}{81}},
  \bibinfo{pages}{41410} (\bibinfo{year}{2010}).

\bibitem[{\citenamefont{Apaja et~al.}(2010)\citenamefont{Apaja, Hyrk\"as, and
  Manninen}}]{Apaja2010}
\bibinfo{author}{\bibfnamefont{V.}~\bibnamefont{Apaja}},
  \bibinfo{author}{\bibfnamefont{M.}~\bibnamefont{Hyrk\"as}}, \bibnamefont{and}
  \bibinfo{author}{\bibfnamefont{M.}~\bibnamefont{Manninen}},
  \bibinfo{journal}{Phys. Rev. A} \textbf{\bibinfo{volume}{82}},
  \bibinfo{pages}{041402} (\bibinfo{year}{2010}).

\bibitem[{\citenamefont{Goldman et~al.}(2011)\citenamefont{Goldman, Urban, and
  Bercioux}}]{Goldman2011}
\bibinfo{author}{\bibfnamefont{N.}~\bibnamefont{Goldman}},
  \bibinfo{author}{\bibfnamefont{D.~F.} \bibnamefont{Urban}}, \bibnamefont{and}
  \bibinfo{author}{\bibfnamefont{D.}~\bibnamefont{Bercioux}},
  \bibinfo{journal}{Phys. Rev. A} \textbf{\bibinfo{volume}{83}},
  \bibinfo{pages}{063601} (\bibinfo{year}{2011}).

\bibitem[{\citenamefont{Vicencio and Mejía-Cortés}(2014)}]{Vicencio2014}
\bibinfo{author}{\bibfnamefont{R.~A.} \bibnamefont{Vicencio}} \bibnamefont{and}
  \bibinfo{author}{\bibfnamefont{C.}~\bibnamefont{Mejía-Cortés}},
  \bibinfo{journal}{Journal of Optics} \textbf{\bibinfo{volume}{16}},
  \bibinfo{pages}{015706} (\bibinfo{year}{2014}).

\bibitem[{\citenamefont{Guzm\'an-Silva
  et~al.}(2014)\citenamefont{Guzm\'an-Silva, Mej\'{\i}a-Cort\'es, Bandres,
  Rechtsman, Weimann, Nolte, Segev, Szameit, and Vicencio}}]{Guzman-Silva2014}
\bibinfo{author}{\bibfnamefont{D.}~\bibnamefont{Guzm\'an-Silva}},
  \bibinfo{author}{\bibfnamefont{C.}~\bibnamefont{Mej\'{\i}a-Cort\'es}},
  \bibinfo{author}{\bibfnamefont{M.~A.} \bibnamefont{Bandres}},
  \bibinfo{author}{\bibfnamefont{M.~C.} \bibnamefont{Rechtsman}},
  \bibinfo{author}{\bibfnamefont{S.}~\bibnamefont{Weimann}},
  \bibinfo{author}{\bibfnamefont{S.}~\bibnamefont{Nolte}},
  \bibinfo{author}{\bibfnamefont{M.}~\bibnamefont{Segev}},
  \bibinfo{author}{\bibfnamefont{A.}~\bibnamefont{Szameit}}, \bibnamefont{and}
  \bibinfo{author}{\bibfnamefont{R.~A.} \bibnamefont{Vicencio}},
  \bibinfo{journal}{New Journal of Physics} \textbf{\bibinfo{volume}{16}},
  \bibinfo{pages}{063061} (\bibinfo{year}{2014}).

\bibitem[{\citenamefont{Bloch et~al.}(2008)\citenamefont{Bloch, Dalibard, and
  Zwerger}}]{Bloch:2008gl}
\bibinfo{author}{\bibfnamefont{I.}~\bibnamefont{Bloch}},
  \bibinfo{author}{\bibfnamefont{J.}~\bibnamefont{Dalibard}}, \bibnamefont{and}
  \bibinfo{author}{\bibfnamefont{W.}~\bibnamefont{Zwerger}},
  \bibinfo{journal}{Reviews of Modern Physics} \textbf{\bibinfo{volume}{80}},
  \bibinfo{pages}{885} (\bibinfo{year}{2008}).

\bibitem[{\citenamefont{Carusotto and Ciuti}(2013)}]{Carusotto:2013gh}
\bibinfo{author}{\bibfnamefont{I.}~\bibnamefont{Carusotto}} \bibnamefont{and}
  \bibinfo{author}{\bibfnamefont{C.}~\bibnamefont{Ciuti}},
  \bibinfo{journal}{Reviews of Modern Physics} \textbf{\bibinfo{volume}{85}},
  \bibinfo{pages}{299} (\bibinfo{year}{2013}).

\bibitem[{\citenamefont{Davis et~al.}(1996)\citenamefont{Davis, Miura,
  Sugimoto, and Hirao}}]{Davis1996}
\bibinfo{author}{\bibfnamefont{K.~M.} \bibnamefont{Davis}},
  \bibinfo{author}{\bibfnamefont{K.}~\bibnamefont{Miura}},
  \bibinfo{author}{\bibfnamefont{N.}~\bibnamefont{Sugimoto}}, \bibnamefont{and}
  \bibinfo{author}{\bibfnamefont{K.}~\bibnamefont{Hirao}},
  \bibinfo{journal}{Opt. Lett.} \textbf{\bibinfo{volume}{21}},
  \bibinfo{pages}{1729} (\bibinfo{year}{1996}).

\bibitem[{\citenamefont{Pertsch et~al.}(1999)\citenamefont{Pertsch, Dannberg,
  Elflein, Br\"auer, and Lederer}}]{Pertsch1999}
\bibinfo{author}{\bibfnamefont{T.}~\bibnamefont{Pertsch}},
  \bibinfo{author}{\bibfnamefont{P.}~\bibnamefont{Dannberg}},
  \bibinfo{author}{\bibfnamefont{W.}~\bibnamefont{Elflein}},
  \bibinfo{author}{\bibfnamefont{A.}~\bibnamefont{Br\"auer}}, \bibnamefont{and}
  \bibinfo{author}{\bibfnamefont{F.}~\bibnamefont{Lederer}},
  \bibinfo{journal}{Phys. Rev. Lett.} \textbf{\bibinfo{volume}{83}},
  \bibinfo{pages}{4752} (\bibinfo{year}{1999}).

\bibitem[{\citenamefont{Morandotti et~al.}(1999)\citenamefont{Morandotti,
  Peschel, Aitchison, Eisenberg, and Silberberg}}]{Morandotti1999}
\bibinfo{author}{\bibfnamefont{R.}~\bibnamefont{Morandotti}},
  \bibinfo{author}{\bibfnamefont{U.}~\bibnamefont{Peschel}},
  \bibinfo{author}{\bibfnamefont{J.~S.} \bibnamefont{Aitchison}},
  \bibinfo{author}{\bibfnamefont{H.~S.} \bibnamefont{Eisenberg}},
  \bibnamefont{and}
  \bibinfo{author}{\bibfnamefont{Y.}~\bibnamefont{Silberberg}},
  \bibinfo{journal}{Phys. Rev. Lett.} \textbf{\bibinfo{volume}{83}},
  \bibinfo{pages}{4756} (\bibinfo{year}{1999}).

\bibitem[{\citenamefont{Lenz et~al.}(1999)\citenamefont{Lenz, Talanina, and
  de~Sterke}}]{Lenz1999}
\bibinfo{author}{\bibfnamefont{G.}~\bibnamefont{Lenz}},
  \bibinfo{author}{\bibfnamefont{I.}~\bibnamefont{Talanina}}, \bibnamefont{and}
  \bibinfo{author}{\bibfnamefont{C.~M.} \bibnamefont{de~Sterke}},
  \bibinfo{journal}{Phys. Rev. Lett.} \textbf{\bibinfo{volume}{83}},
  \bibinfo{pages}{963} (\bibinfo{year}{1999}).

\bibitem[{\citenamefont{Chiodo et~al.}(2006)\citenamefont{Chiodo, Valle,
  Osellame, Longhi, Cerullo, Ramponi, Laporta, and Morgner}}]{Chiodo2006}
\bibinfo{author}{\bibfnamefont{N.}~\bibnamefont{Chiodo}},
  \bibinfo{author}{\bibfnamefont{G.~D.} \bibnamefont{Valle}},
  \bibinfo{author}{\bibfnamefont{R.}~\bibnamefont{Osellame}},
  \bibinfo{author}{\bibfnamefont{S.}~\bibnamefont{Longhi}},
  \bibinfo{author}{\bibfnamefont{G.}~\bibnamefont{Cerullo}},
  \bibinfo{author}{\bibfnamefont{R.}~\bibnamefont{Ramponi}},
  \bibinfo{author}{\bibfnamefont{P.}~\bibnamefont{Laporta}}, \bibnamefont{and}
  \bibinfo{author}{\bibfnamefont{U.}~\bibnamefont{Morgner}},
  \bibinfo{journal}{Opt. Lett.} pp. \bibinfo{pages}{1651--1653}
  (\bibinfo{year}{2006}).

\bibitem[{\citenamefont{Dreisow et~al.}(2008)\citenamefont{Dreisow, Heinrich,
  Szameit, Doering, Nolte, Tuennermann, Fahr, and Lederer}}]{Dreisow2008}
\bibinfo{author}{\bibfnamefont{F.}~\bibnamefont{Dreisow}},
  \bibinfo{author}{\bibfnamefont{M.}~\bibnamefont{Heinrich}},
  \bibinfo{author}{\bibfnamefont{A.}~\bibnamefont{Szameit}},
  \bibinfo{author}{\bibfnamefont{S.}~\bibnamefont{Doering}},
  \bibinfo{author}{\bibfnamefont{S.}~\bibnamefont{Nolte}},
  \bibinfo{author}{\bibfnamefont{A.}~\bibnamefont{Tuennermann}},
  \bibinfo{author}{\bibfnamefont{S.}~\bibnamefont{Fahr}}, \bibnamefont{and}
  \bibinfo{author}{\bibfnamefont{F.}~\bibnamefont{Lederer}},
  \bibinfo{journal}{Opt. Express} \textbf{\bibinfo{volume}{16}},
  \bibinfo{pages}{3474} (\bibinfo{year}{2008}).

\bibitem[{\citenamefont{Dreisow
  et~al.}(2009{\natexlab{a}})\citenamefont{Dreisow, Szameit, Heinrich, Pertsch,
  Nolte, T\"unnermann, and Longhi}}]{Dreisow2009}
\bibinfo{author}{\bibfnamefont{F.}~\bibnamefont{Dreisow}},
  \bibinfo{author}{\bibfnamefont{A.}~\bibnamefont{Szameit}},
  \bibinfo{author}{\bibfnamefont{M.}~\bibnamefont{Heinrich}},
  \bibinfo{author}{\bibfnamefont{T.}~\bibnamefont{Pertsch}},
  \bibinfo{author}{\bibfnamefont{S.}~\bibnamefont{Nolte}},
  \bibinfo{author}{\bibfnamefont{A.}~\bibnamefont{T\"unnermann}},
  \bibnamefont{and} \bibinfo{author}{\bibfnamefont{S.}~\bibnamefont{Longhi}},
  \bibinfo{journal}{Phys. Rev. Lett.} \textbf{\bibinfo{volume}{102}},
  \bibinfo{pages}{076802} (\bibinfo{year}{2009}{\natexlab{a}}).

\bibitem[{\citenamefont{Dreisow
  et~al.}(2009{\natexlab{b}})\citenamefont{Dreisow, Szameit, Heinrich, Nolte,
  T\"unnermann, Ornigotti, and Longhi}}]{2Dreisow2009}
\bibinfo{author}{\bibfnamefont{F.}~\bibnamefont{Dreisow}},
  \bibinfo{author}{\bibfnamefont{A.}~\bibnamefont{Szameit}},
  \bibinfo{author}{\bibfnamefont{M.}~\bibnamefont{Heinrich}},
  \bibinfo{author}{\bibfnamefont{S.}~\bibnamefont{Nolte}},
  \bibinfo{author}{\bibfnamefont{A.}~\bibnamefont{T\"unnermann}},
  \bibinfo{author}{\bibfnamefont{M.}~\bibnamefont{Ornigotti}},
  \bibnamefont{and} \bibinfo{author}{\bibfnamefont{S.}~\bibnamefont{Longhi}},
  \bibinfo{journal}{Phys. Rev. A} \textbf{\bibinfo{volume}{79}},
  \bibinfo{pages}{055802} (\bibinfo{year}{2009}{\natexlab{b}}).

\bibitem[{\citenamefont{URL}()}]{sup}
\bibinfo{author}{\bibnamefont{URL}}, \bibinfo{note}{to be provided by Editor}.

\bibitem[{\citenamefont{Ams et~al.}(2005)\citenamefont{Ams, Marshall, Spence,
  and Withford}}]{Ams2005}
\bibinfo{author}{\bibfnamefont{M.}~\bibnamefont{Ams}},
  \bibinfo{author}{\bibfnamefont{G.}~\bibnamefont{Marshall}},
  \bibinfo{author}{\bibfnamefont{D.}~\bibnamefont{Spence}}, \bibnamefont{and}
  \bibinfo{author}{\bibfnamefont{M.}~\bibnamefont{Withford}},
  \bibinfo{journal}{Opt. Express} \textbf{\bibinfo{volume}{13}},
  \bibinfo{pages}{5676} (\bibinfo{year}{2005}).

\bibitem[{\citenamefont{Cheng et~al.}(2003)\citenamefont{Cheng, Sugioka,
  Midorikawa, Masuda, Toyoda, Kawachi, and Shihoyama}}]{Cheng2003}
\bibinfo{author}{\bibfnamefont{Y.}~\bibnamefont{Cheng}},
  \bibinfo{author}{\bibfnamefont{K.}~\bibnamefont{Sugioka}},
  \bibinfo{author}{\bibfnamefont{K.}~\bibnamefont{Midorikawa}},
  \bibinfo{author}{\bibfnamefont{M.}~\bibnamefont{Masuda}},
  \bibinfo{author}{\bibfnamefont{K.}~\bibnamefont{Toyoda}},
  \bibinfo{author}{\bibfnamefont{M.}~\bibnamefont{Kawachi}}, \bibnamefont{and}
  \bibinfo{author}{\bibfnamefont{K.}~\bibnamefont{Shihoyama}},
  \bibinfo{journal}{Opt. Lett.} \textbf{\bibinfo{volume}{28}},
  \bibinfo{pages}{55} (\bibinfo{year}{2003}).

\bibitem[{\citenamefont{Salter et~al.}(2014)\citenamefont{Salter, Baum,
  Alexeev, Schmidt, and Booth}}]{Salter2014}
\bibinfo{author}{\bibfnamefont{P.~S.} \bibnamefont{Salter}},
  \bibinfo{author}{\bibfnamefont{M.}~\bibnamefont{Baum}},
  \bibinfo{author}{\bibfnamefont{I.}~\bibnamefont{Alexeev}},
  \bibinfo{author}{\bibfnamefont{M.}~\bibnamefont{Schmidt}}, \bibnamefont{and}
  \bibinfo{author}{\bibfnamefont{M.~J.} \bibnamefont{Booth}},
  \bibinfo{journal}{Opt. Express} \textbf{\bibinfo{volume}{22}},
  \bibinfo{pages}{17644} (\bibinfo{year}{2014}).

\bibitem[{\citenamefont{Eaton et~al.}(2009)\citenamefont{Eaton, Chen, Zhang,
  Iyer, Li, Ng, Ho, Aitchison, and Herman}}]{Eaton2009}
\bibinfo{author}{\bibfnamefont{S.~M.} \bibnamefont{Eaton}},
  \bibinfo{author}{\bibfnamefont{W.-J.} \bibnamefont{Chen}},
  \bibinfo{author}{\bibfnamefont{H.}~\bibnamefont{Zhang}},
  \bibinfo{author}{\bibfnamefont{R.}~\bibnamefont{Iyer}},
  \bibinfo{author}{\bibfnamefont{J.}~\bibnamefont{Li}},
  \bibinfo{author}{\bibfnamefont{M.}~\bibnamefont{Ng}},
  \bibinfo{author}{\bibfnamefont{S.}~\bibnamefont{Ho}},
  \bibinfo{author}{\bibfnamefont{J.}~\bibnamefont{Aitchison}},
  \bibnamefont{and} \bibinfo{author}{\bibfnamefont{P.~R.}
  \bibnamefont{Herman}}, \bibinfo{journal}{J. Lightwave Technol.}
  \textbf{\bibinfo{volume}{27}}, \bibinfo{pages}{1079} (\bibinfo{year}{2009}).

\bibitem[{\citenamefont{Umucalılar and Carusotto}(2012)}]{Umucallar:2012bo}
\bibinfo{author}{\bibfnamefont{R.~O.} \bibnamefont{Umucalılar}}
  \bibnamefont{and}
  \bibinfo{author}{\bibfnamefont{I.}~\bibnamefont{Carusotto}},
  \bibinfo{journal}{Phys. Rev. Lett.} \textbf{\bibinfo{volume}{108}},
  \bibinfo{pages}{206809} (\bibinfo{year}{2012}).

\bibitem[{\citenamefont{Maghrebi et~al.}(2014)\citenamefont{Maghrebi, Yao,
  Hafezi, Pohl, Firstenberg, and Gorshkov}}]{Maghrebi:2014tq}
\bibinfo{author}{\bibfnamefont{M.~F.} \bibnamefont{Maghrebi}},
  \bibinfo{author}{\bibfnamefont{N.~Y.} \bibnamefont{Yao}},
  \bibinfo{author}{\bibfnamefont{M.}~\bibnamefont{Hafezi}},
  \bibinfo{author}{\bibfnamefont{T.}~\bibnamefont{Pohl}},
  \bibinfo{author}{\bibfnamefont{O.}~\bibnamefont{Firstenberg}},
  \bibnamefont{and} \bibinfo{author}{\bibfnamefont{A.~V.}
  \bibnamefont{Gorshkov}}, 
  \bibinfo{note}{arXiv:1411.6624}.

\bibitem[{\citenamefont{Vicencio et~al.}()\citenamefont{Vicencio, Cantillano,
  Morales-Inostroza, Real, Weimann, Szameit, and Molina}}]{Vicencio2014b}
\bibinfo{author}{\bibfnamefont{R.}~\bibnamefont{Vicencio}},
  \bibinfo{author}{\bibfnamefont{C.}~\bibnamefont{Cantillano}},
  \bibinfo{author}{\bibfnamefont{L.}~\bibnamefont{Morales-Inostroza}},
  \bibinfo{author}{\bibfnamefont{B.}~\bibnamefont{Real}},
  \bibinfo{author}{\bibfnamefont{S.}~\bibnamefont{Weimann}},
  \bibinfo{author}{\bibfnamefont{A.}~\bibnamefont{Szameit}}, \bibnamefont{and}
  \bibinfo{author}{\bibfnamefont{M.}~\bibnamefont{Molina}},
  \bibinfo{note}{arXiv:1412.4713}.

\end{thebibliography}
\end{document}